\documentclass[12pt]{article}

\usepackage[height=8.5in,width=6.4in]{geometry}

\usepackage{xparse}
\usepackage{amssymb, amsmath}
\usepackage{xcolor}
\usepackage{tikz}
\usepackage{cite}
\usepackage[linktocpage]{hyperref}
\usepackage{mathtools}
\usepackage[vcentermath]{youngtab}

\DeclareFontShape{OT1}{cmr}{mx}{n}%
    {<->cmr10}{}

\renewcommand{\hat}{\widehat}
\renewcommand{\tilde}{\widetilde}

\numberwithin{equation}{section}

\begin{document}
\setcounter{tocdepth}{2}

\begin{titlepage}
\begin{flushright} 
NITEP 140\\
OCU-PHYS 567
\end{flushright}

\vskip .8cm

\begin{center}

{\LARGE \fontseries{mx}\selectfont
On the Nekrasov Partition Function of Gauged Argyres-Douglas Theories
\par
}

\vskip 1.2cm

Takuya Kimura,$^{\diamondsuit, 1}$ and Takahiro Nishinaka,$^{\clubsuit, 2,3,4}$

\vskip .8cm

{\it
$^1$ Department of Physical Sciences, College of Science and Engineering\\
 Ritsumeikan University, Shiga 525-8577, Japan
}

\vskip.4cm

{\it
$^2$ Department of Physics, Graduate School of Science\\
Osaka Metropolitan University, Osaka 558-8585, Japan
}

\vskip.4cm

{\it
$^3$ Nambu Yoichiro Institute of Theoretical and Experimental Physics (NITEP)\\
Osaka Metropolitan University, Osaka 558-8585, Japan
}

\vskip.4cm

{\it
$^4$ Osaka Central Advanced Mathematical Institute (OCAMI)\\
Osaka Metropolitan University, Osaka 558-8585, Japan
}

\end{center}

\vskip.5cm

\begin{abstract}
We study $SU(2)$ gauge theories coupled to $(A_1,D_N)$ theories with
 or without a fundamental hypermultiplet. For even $N$, a formula for
 the contribution of $(A_1,D_N)$ to the Nekrasov partition function was recently
 obtained by us with Y.~Sugawara and T.~Uetoko. In this paper, we
 generalize it to
 the case of odd $N$ in the classical limit, under the condition that the relevant couplings
 and vacuum expectation values of Coulomb branch operators of
 $(A_1,D_N)$ are all turned off. We apply our formula to the $(A_2,A_5)$
 theory to find that its prepotential is related to that of the $SU(2)$
 gauge theory with four fundamental flavors by a simple change of
 variables.

\end{abstract}

\renewcommand{\thefootnote}{\fnsymbol{footnote}}
\footnotetext[0]{
\!\!\!\!\!\!\!\!\!\!\!$^{\diamondsuit}$rp0047ir@ed.ritsumei.ac.jp,
\\
$^{\clubsuit}$nishinaka@osaka-cu.ac.jp}
\renewcommand{\thefootnote}{\arabic{footnote}}

\end{titlepage}

\tableofcontents

\section{Introduction}

The Nekrasov's instanton partition function \cite{Nekrasov:2002qd} for 4D $\mathcal{N}=2$ gauge theories
has uncovered various non-perturbative phenomena in these
theories. For instance, the Seiberg-Witten prepotential was
 derived from the path integral \cite{Nekrasov:2002qd, Nekrasov:2003rj},
a relation to
integrable systems was discovered \cite{Nekrasov:2009rc}, and a novel 2d/4d
correspondence called the AGT correspondence was discovered
\cite{Alday:2009aq, Wyllard:2009hg}.

A generalization of the above success to theories
coupled to a
strongly-coupled superconformal field theories (SCFTs) has partially been
studied. In particular, the AGT correspondence has been generalized in
\cite{Bonelli:2011aa, Gaiotto:2012sf} to gauge theories coupled to
Argyres-Douglas (AD) theories. We call these gauge theories ``gauged AD
theories.'' Since
AD theories have no weak-coupling limit, supersymmetric localization is not available for these
theories. As a result, the generalized AGT correspondence has been
the only promising way of evaluating the
instanton partition function of these theories.

One restriction of the generalized AGT correspondence was, however, that
it
was only applied to non-conformally gauged AD
theories.\footnote{Here, by ``non-conformally gauged,'' we mean that the
beta function of the gauge coupling is asymptotically free.} The reason
for this is that conformally gauged AD theories have no known
realization from 6d (2,0) $A_1$ theory, and therefore the AGT
correspondence is not directly applied to them. As a result, until recently, the instanton
partition function of conformally gauged AD theories was not evaluated.

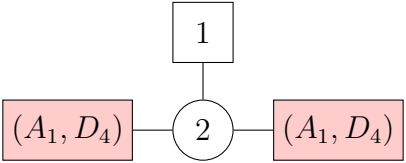
\begin{figure}
    \begin{center}
   \begin{tikzpicture}[gauge/.style={circle,draw=black,inner sep=0pt,minimum size=8mm},flavor/.style={rectangle,draw=black,inner sep=0pt,minimum size=8mm},AD/.style={rectangle,draw=black,fill=red!20,inner sep=0pt,minimum size=8mm},auto]
   \node[AD] (0) at (-1.8,0) {\;$(A_1,D_{4})$\;};
   \node[gauge] (1) at (0,0) [shape=circle] {\;$2$\;} edge (0);
   \node[AD] (2) at (1.8,0)  {\;$(A_1,D_{4})$\;} edge (1);
   \node[flavor] (3) at (0,1.3) {\;1\;} edge (1);
     \end{tikzpicture}
   \caption{The $(A_3,A_3)$ theory is
     identical to the conformal $SU(2)$ gauge theory coupled to two
     $(A_1,D_4)$ theories and a fundamental hypermultiplet of
     $SU(2)$. Here, the middle circle with $2$ inside stands for an
     $SU(2)$ vector multiplet, and the top box with $1$ inside stands
     for a fundamental hypermultiplet.}
   \label{fig:quiver1}
    \end{center}
\end{figure}

A first idea of computing the instanton partition function of
conformally gauged AD theories has been
provided in \cite{Kimura:2020krd}. A key ingredient is the $U(2)$ version of the generalized AGT
correspondence, which is stated in terms of irregular states of the direct sum of Virasoro and Heisenberg
algebra $Vir\oplus H$. 
For instance,
 let us consider $SU(2)$ gauge theory coupled to a fundamental
 hypermultiplet and two copies of AD theory called $(A_1,D_4)$ (Fig.~\ref{fig:quiver1}). Here, the ``matter'' sector is precisely chosen so
 that the beta function of the $SU(2)$ gauge coupling
 vanishes. This coupled theory is also known as the
 ``$(A_3,A_3)$ theory.'' While the AGT correspondence cannot be directly
 applied to the $(A_3,A_3)$ theory,
one can apply it to a factor in the following decomposition
of the partition function:
\begin{align}
 \mathcal{Z} =
 \mathcal{Z}_\text{pert}\sum_{Y_1,Y_2}q^{|Y_1|+|Y_2|}\mathcal{Z}^\text{vec}_{Y_1,Y_2}(a)\mathcal{Z}^\text{fund}_{Y_1,Y_2}(a,M)\prod_{i=1}^2\mathcal{Z}^{(A_1,D_{4})}_{Y_1,Y_2}(a,m_i,d_i,u_i)~,
\label{eq:U2-1}
\end{align}
where $a$ is the vacuum expectation value (VEV) of the Coulomb branch
operator in the vector multiplet, $q$ is the exponential of
the gauge coupling, the sum runs over pairs of Young
diagrams $(Y_1,Y_2)$, $|Y|$ stands for the number of boxes in a Young
diagram $Y$,
and $\mathcal{Z}^\text{vec}_{Y_1,Y_2}$ and
$\mathcal{Z}^\text{fund}_{Y_1,Y_2}$ are the contributions from the vector and hypermultiplets.\footnote{Here $\mathcal{Z}_\text{pert}$ is a prefactor that makes the
$q$-series start with $1$.} 
The factor
$\mathcal{Z}_{Y_1,Y_2}^{(A_1,D_4)}$ is the contribution from an
$(A_1,D_4)$ theory, which is hard to evaluate via localization but can
be evaluated via the $U(2)$-version of the generalized AGT
correspondence \cite{Kimura:2020krd}.

The reason for the ``$U(2)$-version'' is that the decomposition
\eqref{eq:U2-1} is possible only when the gauge group is $U(2)$ instead
of $SU(2)$.
 The difference between $U(2)$ and
$SU(2)$ gives rise to a prefactor of the partition function, known as the
$U(1)$-factor.  By factoring out the $U(1)$-factor, one can read off the
partition function and the prepotential of
the $(A_3,A_3)$ theory from \eqref{eq:U2-1}.  As discussed in \cite{Kimura:2020krd}, when dimensionful parameters are turned off except for $a$, the prepotential
$\mathcal{F}_{(A_3,A_3)}(q;a)$ of the $(A_3,A_3)$ theory read off as
above is in a surprising relation to the prepotential
$\mathcal{F}_{SU(2)}^{N_f=4}(q;a)$ of $SU(2)$ gauge theory with four
fundamental flavors, i.e.,
\begin{align}
 2\mathcal{F}_{(A_3,A_3)}(q;a) = \mathcal{F}_{SU(2)}^{N_f=4}(q^2;a)~.
\label{eq:rel1}
\end{align}
This remarkable relation was then used to read off how the S-duality of
$(A_3,A_3)$ acts on the UV gauge coupling $q$.

While the above $U(2)$-version of the generalized AGT correspondence provides a novel way of
evaluating the instanton partition function of conformally gauged AD theories, one of its restrictions is that the formula
provided in \cite{Kimura:2020krd} is only for $(A_1,D_\text{even})$ theories. The
reason for this is that only irregular states of {\it integer}
ranks were constructed in \cite{Kimura:2020krd}, and those of {\it half-integer} ranks
are still to be identified.\footnote{As explained in the next section,
the rank of an irregular state $|I\rangle$ is defined by the maximal
$n\in \mathbb{N}/2$ such that $L_{2n}|I\rangle \neq 0$.}

In this paper, we extend the result of \cite{Kimura:2020krd} to the case of
$(A_1,D_\text{odd})$ theories, under the condition that all couplings
and VEVs of Coulomb branch operators in $(A_1,D_\text{odd})$ are turned off. This is done by explicitly
 identifying the action of $Vir\times H$ on irregular states of
 half-integer ranks. This action turns out to be very simple in the
 classical limit $\epsilon_1,\epsilon_2\to 0$, when the above condition
 is satisfied.

\begin{figure}
    \begin{center}
   \begin{tikzpicture}[gauge/.style={circle,draw=black,inner sep=0pt,minimum size=8mm},flavor/.style={rectangle,draw=black,inner sep=0pt,minimum size=8mm},AD/.style={rectangle,draw=black,fill=red!20,inner sep=0pt,minimum size=8mm},auto]

   \node[AD] (1) at (-1.8,0) {\;$(A_1,D_{3})$\;};
   \node[gauge] (2) at (0,0) [shape=circle] {\;$2$\;} edge (1);
   \node[AD] (3) at (1.8,0)  {\;$(A_1,D_{6})$\;} edge (2);
   \node[flavor] (4) at (0,1.3) {\;1\;} edge (2);
   
     \end{tikzpicture}
   \caption{The $(A_2,A_5)$ theory is an $\mathcal{N}=2$ SCFT, which is identical to a conformal $SU(2)$
     gauging of $(A_1,D_3)$ and $(A_1,D_6)$ theories together with a
     fundamental hypermultiplet of $SU(2)$.}
   \label{fig:quiver2}
    \end{center}
   \end{figure}
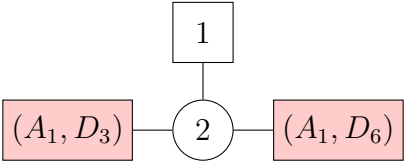

As an application of our extension, we evaluate the prepotential
of the
 $(A_2,A_5)$ theory, which is the conformal $SU(2)$ gauge theory coupled
 to a fundamental hypermultiplet and AD theories called $(A_1,D_6)$ and
 $(A_1,D_3)$ (Fig.~\ref{fig:quiver2}).\footnote{See
 \cite{Giacomelli:2020ryy} for a recent discussion on
 the conformal manifold of $(A_n, A_m)$ theories.} To compute the
 partition function $\mathcal{Z}_{(A_2,A_5)}$ of this theory, one needs to know the contribution
 of the $(A_1,D_3)$ and $(A_1,D_6)$ theories at each fixed point on the instanton moduli
 space, i.e.~, $\mathcal{Z}_{Y_1,Y_2}^{(A_1,D_3)}$
 and $\mathcal{Z}_{Y_1,Y_2}^{(A_1,D_6)}$. While the latter can be
 evaluated via the method of \cite{Kimura:2020krd}, computing the former
 needs a prescription that we develop in this paper. We then read off
 from $\mathcal{Z}_{(A_2,A_5)}$ an expression for the
prepotential $\mathcal{F}_{(A_2,A_5)}$ of the $(A_2,A_5)$ theory, which
turns out to be in a
surprising relation to
$\mathcal{F}_{SU(2)}^{N_f=4}$:
\begin{align}
 3\mathcal{F}_{(A_2,A_5)}(q;a) = \mathcal{F}_{SU(2)}^{N_f=4}(q^3;a)~.
\end{align}
Note that this relation is quite similar to \eqref{eq:rel1} but different. From
this relation, we read off how the S-duality group acts on the UV gauge
coupling $q$ of the $(A_2,A_5)$ theory. 
A generalization of our result to the case of
 all dimensionful parameters turned on is left for future work.

The organization of this paper is the following. In
Sec.~\ref{sec:review}, we review the generalized AGT correspondence and
its $U(2)$-version. In Sec.~\ref{sec:U2}, we consider the generalization
of the $U(2)$-version to $(A_1,D_\text{odd})$. In Sec.~\ref{sec:A2A5},
we apply a formula developed in Sec.~\ref{sec:U2} to the $(A_2,A_5)$
theory and show that the prepotential of $(A_2,A_5)$ is related to that
of $SU(2)$ superconformal QCD by a change of variables. In
Sec.~\ref{sec:SW}, we show that the prepotential relation found in
Sec.~\ref{sec:A2A5} is consistent with the Seiberg-Witten curve.

\section{$U(2)$-version of generalized AGT for $(A_1,D_\text{even})$}
\label{sec:review}

In this section, we give a brief review of the $U(2)$-version of the
generalized AGT correspondence for $(A_1,D_{N})$ theories with even $N$.

\subsection{Generalized AGT correspondence}
\label{subsec:GAGT}

We first review the original generalized AGT correspondence.
Recall that the $(A_1,D_{N})$ theory for a positive integer $N\geq 2$ is realized by compactifying the 6d
(2,0) $A_1$ theory on a sphere with two punctures, one of which is a
regular puncture and the other is an irregular puncture of rank
$N/2$ \cite{Bonelli:2011aa,Gaiotto:2012sf, Xie:2012hs}. These punctures specify how the Higgs field $\Phi(z)$ in the
corresponding Hitchin system behaves
around them; $\Phi(z)$ has a simple pole at a regular puncture
while it behaves as $\Phi(z) \sim 1/z^{N/2+1}$ around an irregular
puncture of rank $N/2$, where we take $z=0$ as the
locus of the puncture.

According to the generalized AGT correspondence \cite{Bonelli:2011aa, Gaiotto:2012sf}, the regular puncture
corresponds to a Virasoro primary state $|a\rangle$, and the
irregular puncture corresponds to an irregular state $|I^{(N/2)}\rangle$ of
Virasoro algebra at central charge $c=1+6Q^2$.
While there are two different characterizations of $|I^{(N/2)}\rangle$, we will use the one discussed in \cite{Gaiotto:2012sf}.
 Here, the irregular state
is not a primary state but
a simultaneous eigen state of $L_k$ for $k\geq \lceil N/2
\rceil$, with vanishing eigenvalues for $k> N$. 
Therefore, an irregular state $|I^{(N/2)}\rangle$ satisfies 
\begin{align}
 L_k|I^{(N/2)}\rangle = 
\left\{
\begin{array}{l}
0 \qquad \text{for}\qquad N<k
\\[2mm]
\lambda_k|I^{(N/2)}\rangle\qquad
 \text{for}\qquad \left\lceil \frac{N}{2}
 \right\rceil\leq k\leq  N\\
\end{array}
\right.~,
\label{eq:eigen1}
\end{align}
for a set of eigenvalues $\{\lambda_{\lceil
N/2\rceil},\cdots,\lambda_N\}$.\footnote{While $\lceil
N/2\rceil = N/2$ for even $N$, we here write things so that they can be
easily generalized to odd $N$ in the next section.}
This characterization of the irregular
state is such that
\begin{align}
 x^2 = -\frac{\langle a | T(z)|I^{(N/2)}\rangle}{\langle
 a|I^{(N/2)}\rangle}
\label{eq:curve0}
\end{align}
is equivalent to the Seiberg Witten (SW) curve of the 4d theory. Indeed,
from \eqref{eq:eigen1}, we see that \eqref{eq:curve0} is evaluated as
\begin{align}
 x^2 = -\frac{\lambda_N}{z^{N+2}} - \frac{\lambda_{N-1}}{z^{N+1}} - \cdots  - \frac{a(Q-a)}{z^2}~,
\label{eq:curve1}
\end{align}
which is identical to the SW curve of the $(A_1,D_N)$ theory. 

Given the above regular state $|a\rangle$ and the irregular state
$|I^{(N/2)}\rangle$, the generalized AGT correspondence states that
\begin{align}
 \mathcal{Z}_{(A_1,D_N)} = \langle a |I^{(N/2)}\rangle~,
\label{eq:ADN}
\end{align}
is identified with the Nekrasov partition function of the $(A_1,D_N)$
theory. Note here that, since no weakly-coupled description is known for
this theory, the above partition function cannot be evaluated by
supersymmetric localization.

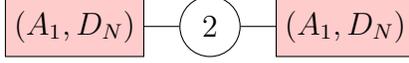
\begin{figure}
    \begin{center}
   \begin{tikzpicture}[gauge/.style={circle,draw=black,inner sep=0pt,minimum size=8mm},flavor/.style={rectangle,draw=black,inner sep=0pt,minimum size=8mm},AD/.style={rectangle,draw=black,fill=red!20,inner sep=0pt,minimum size=8mm},auto]

   \node[AD] (1) at (-1.8,0) {\;$(A_1,D_{N})$\;};
   \node[gauge] (2) at (0,0) [shape=circle] {\;$2$\;} edge (1);
   \node[AD] (3) at (1.8,0)  {\;$(A_1,D_{N})$\;} edge (2);
   
     \end{tikzpicture}
   \caption{$SU(2)$ gauge theory coupled to two $(A_1,D_N)$ theories}
   \label{fig:quiver4}
    \end{center}
   \end{figure}

Similarly, 4d $\mathcal{N}=2$ $SU(2)$ gauge theory coupled to two copies
 of $(A_1,D_N)$ (Fig.~\ref{fig:quiver4}) is constructed by compactifying 6d (2,0) $A_1$ theory on
 sphere with two irregular singularities of rank $N/2$. The generalized
 AGT correspondence then implies that the Nekrasov partition function of this
 theory is given by
\begin{align}
 \mathcal{Z}_{SU(2)}^{2\times (A_1,D_N)} = \langle I^{(N/2)}|
 I^{(N/2)}\rangle~.
\label{eq:AD-MN}
\end{align}

Note here that the characterization \eqref{eq:eigen1} does not fix the irregular state $|I^{(N/2)}\rangle$. In
particular, the actions of $L_0,\cdots,L_{\lfloor N/2\rfloor}$ are not
specified there. When $N$ is even, these
actions are expressed in terms of differential operators with respect to
 $(N/2 +1)$ parameters, $c_0,\cdots,c_{N/2}$ \cite{Gaiotto:2012sf}:
\begin{align}
    L_k|I^{(N/2)}\rangle = \left\{
    \begin{array}{l}
     0 \qquad \text{for}\qquad N<k
     \\[2mm]
    \lambda_k|I^{(N/2)}\rangle \qquad \text{for} \quad \frac{N}{2}\leq k\leq N
    \\[2mm]
    \left(\lambda_k + \sum_{\ell = 1}^{N/2-k}\ell\, c_{\ell+k} \frac{\partial
     }{\partial c_{\ell}}\right)|I^{(N/2)}\rangle \qquad \text{for} \quad
    0\leq k < \frac{N}{2}
    \\
    \end{array}
    \right.~,
    \label{eq:eigen2}
    \end{align}
    where the non-vanishing eigenvalues, $\lambda_k$, of $L_{N/2},\cdots,L_N$ are fixed by $c_0,\cdots,c_{N/2}$  
    as
\begin{align}
    \lambda_k = 
    \left\{
    \begin{array}{l}
    -\sum_{\ell=k-N/2}^{N/2} c_\ell c_{k-\ell} \quad \text{for}\quad \frac{N}{2}<k\leq N
    \\[2mm]
    -\sum_{\ell=0}^{k} c_\ell c_{k-\ell} + (k+1)Qc_k \quad \text{for}\quad k\leq \frac{N}{2}
    \end{array}
    \right.~.
    \label{eq:lambda}
\end{align}
The above actions of $L_0,\cdots,L_{N/2}$ follow from the
construction of $|I^{(N/2)}\rangle$ for even $N$ in a colliding limit of
regular primary operators.
Similar colliding limit is not known for odd $N$, and therefore the
actions of $L_0,\cdots,L_{\frac{N-1}{2}}$ have not been identified in
the literature..\footnote{There is another characterization of
the irregular state $|I^{(N/2)}\rangle$ \cite{Bonelli:2011aa}, where
$|I^{(N/2)}\rangle$ has an explicit expression and is an eigen state of
$L_N$ and $L_1$ for even and odd $N$. In this paper, we use the one discussed in
\cite{Gaiotto:2012sf} since it can easily be extended to the $U(2)$-version
that we will review in the next sub-section. It would be an interesting
open problem to consider the $U(2)$-version of the one discussed in \cite{Bonelli:2011aa}.}

\subsection{$U(2)$-version for even $N$}
\label{subsec:U2}

In this sub-section, we discuss the $U(2)$-version of the generalized AGT
correspondence. Here, we focus on irregular states $|I^{(N/2)}\rangle$
for even $N$, and
therefore on $(A_1,D_\text{even})$ theories.

Such a $U(2)$-version was considered in \cite{Kimura:2020krd} in order to
compute the instanton partition function of the $(A_3,A_3)$
theory. Here, the $(A_3,A_3)$ theory is an $\mathcal{N}=2$ superconformal $SU(2)$ gauge theory coupled
to two $(A_1,D_4)$ theories and a fundamental hypermultiplet (Fig.~\ref{fig:quiver1} ).
When the fundamental hypermultiplet is absent, one can compute the
partition function  via the generalized AGT correspondence as in
\eqref{eq:AD-MN}, but its generalization to the $(A_3,A_3)$ theory is
not straightforward. The reason for this is that $(A_3,A_3)$ has no known realization from 6d
(2,0) $A_1$ theory.

Therefore, a more indirect route was taken in \cite{Kimura:2020krd} to compute the
partition function of the $(A_3,A_3)$ theory. First, the generalized AGT
correspondence was extended to the case of $U(2)$ gauge
group. Corresponding to the extra $U(1)$ part of the gauge group, the
Virasoro algebra on the 2d side is now accompanied with an extra Heisenberg
algebra \cite{Alba:2010qc}. The Virasoro irregular state $|I^{(N/2)}\rangle$ is then promoted to an irregular state $|\hat{I}^{(N/2)}\rangle$
of the direct sum of 
Virasoro and Heisenberg algebras $Vir\oplus H$, which  This state is generally
decomposed as 
\begin{align}
 |\hat{I}^{(N/2)}\rangle = |I^{(N/2)}\rangle \otimes |I^{(N/2)}_H\rangle~,
\end{align}
where $|I^{(N/2)}\rangle$ is the Virasoro irregular state satisfying \eqref{eq:eigen2}, and $|I^{(N/2)}_H\rangle$ is an
irregular state of the Heisenberg algebra characterized by
\begin{align}
a_k|I^{(N/2)}_H\rangle = \left\{
\begin{array}{l}
0 \qquad \text{for}\qquad N/2<k~,
\\[2mm]
-ic_k|I^{(N/2)}\rangle\qquad \text{for}\qquad 1\leq k\leq n~.
\end{array}
\right.~,
\label{eq:irreg-H}
\end{align}
where $a_k$ is the basis of the Heisenberg algebra such that
$\{a_k,a_\ell\} = \frac{k}{2}\delta_{k+\ell,0}$~. 
Given the irregular state $|\hat{I}^{(N/2)}\rangle$ of $Vir \oplus
H$, the partition function of $U(2)$ gauge theory coupled to two $(A_1,D_N)$ theories is identified as
\begin{align}
 \mathcal{Z}_{U(2)}^{2\times (A_1,D_N)} = \langle
 \hat{I}^{(N/2)}|\hat{I}^{(N/2)}\rangle~,
\label{eq:U2}
\end{align}
which is a natural generalization of \eqref{eq:AD-MN} to the $U(2)$
gauge group. 

A nice feature of this generalization is that the
highest-weight module of $Vir\oplus H$ has an orthogonal basis
$|a;Y_1,Y_2\rangle$ labeled by two Young diagrams, $Y_1$ and $Y_2$,
that satisfies \cite{Alba:2010qc}
\begin{align}
 {\bf 1} =
 \sum_{Y_1,Y_2}\mathcal{Z}_{Y_1,Y_2}^\text{vec}(a)\,|a;Y_1,Y_2\rangle
 \langle a;Y_1,Y_2|~,
\label{eq:decomp}
\end{align}
where $\mathcal{Z}_{Y_1,Y_2}^\text{vec}(a)$ is the contribution from a
$U(2)$ vector multiplet to the Nekrasov partition function, at the fixed
point corresponding to $(Y_1,Y_2)$ on the moduli space of $U(2)$ instantons. Here
$|a;Y_1,Y_2\rangle$ is a linear combination of states of the form
$L_{-n_1}^{p_1}\cdots L_{-n_k}^{p_k} a_{-m_1}^{q_1}\cdots
a_{-m_\ell}^{q_\ell}|a\rangle$, and $\langle a;Y_1,Y_2|$ is obtained by
replacing each of these states with $\langle a |a_{m_\ell}^{q_{\ell}}
\cdots a_{1}^{q_1} L_{n_k}^{p_k}\cdots L_{n_1}^{p_1}$ {\it without
changing the coefficients of the linear combination.}

One
can use \eqref{eq:decomp} to decompose \eqref{eq:U2} as
\begin{align}
 \mathcal{Z}_{U(2)}^{2\times (A_1,D_N)} =
 \mathcal{Z}_\text{pert}\sum_{Y_1,Y_2}\Lambda^{b_0(|Y_1|+|Y_2|)}\mathcal{Z}_{Y_1,Y_2}^\text{vec}(a)
 \mathcal{Z}_{Y_1,Y_2}^{(A_1,D_N)}(a,m,\pmb{d},\pmb{u})\tilde{\mathcal{Z}}_{Y_1,Y_2}^{(A_1,D_N)}(a,\tilde{m},\tilde{\pmb{d}},\tilde{\pmb{u}})~,
\label{eq:U2A1DN}
\end{align}
where $\mathcal{Z}_\text{pert} \equiv \langle
\hat{I}^{(N/2)}|a\rangle\langle a|\hat{I}^{(N/2)}\rangle,\, \Lambda
\equiv -\zeta^2c_{N/2}\tilde{c}^{*}_{N/2}$ and 
\begin{align}
 \mathcal{Z}_{Y_1,Y_2}^{(A_1,D_N)}(a,m,\pmb{d},\pmb{u}) &\equiv (\zeta c_{N/2})^{-\frac{2(|Y_1|+|Y_2|)}{N}}
 \frac{\langle a;Y_1,Y_2|\hat{I}^{(N/2)}\rangle}{\langle a|
 \hat{I}^{(N/2)}\rangle}~,
\label{eq:AD1}
\\[2mm]
\tilde{\mathcal{Z}}_{Y_1,Y_2}^{(A_1,D_N)}(a,\tilde{m},\tilde{\pmb{d}},\tilde{\pmb{u}}) &\equiv (-\zeta \tilde{c}^{*}_{N/2})^{-\frac{2(|Y_1|+|Y_2|)}{N}}
 \frac{\langle \hat{I}^{(N/2)}|a;Y_1,Y_2\rangle}{\langle\hat{I}^{(N/2)}|
 a\rangle}~.
\label{eq:AD2}
\end{align}
Here, $m,\,\pmb{d}\equiv(d_1,\cdots, d_{\frac{N}{2}-1})$ and $\pmb{u} \equiv
(u_1,\cdots,u_{\frac{N}{2}-1})$ are respectively, a mass parameter, relevant couplings and the VEVs of Coulomb branch
operators. These are related to two-dimensional parameters by
\begin{align}
 d_k &=
 \sum_{\ell=\frac{N}{2}-k}^{\frac{N}{2}}\frac{c_{\ell}c_{N-k-\ell}}{(c_{\frac{N}{2}})^{2-\frac{2k}{N}}}~,\qquad
 m =
 \sum_{\ell=0}^{\frac{N}{2}}\frac{c_{\ell}c_{\frac{N}{2}-\ell}}{c_{\frac{N}{2}}}~,
\label{eq:dk}
\\
 u_k &=
 \sum_{\ell=0}^{\frac{N}{2}-k}\frac{c_{\ell}c_{\frac{N}{2}-k-\ell}}{(c_{\frac{N}{2}})^{1-\frac{2k}{N}}}-
 \sum_{\ell=1}^k \ell
 \frac{c_{\frac{N}{2}+\ell-k}}{(c_{\frac{N}{2}})^{1-\frac{2k}{N}}}\frac{\partial
 \mathcal{F}_{(A_1,D_N)}}{\partial
 c_\ell}~,
\label{eq:uk}
\end{align}
where $\mathcal{F}_{(A_1,D_N)} \equiv \lim_{\epsilon_i\to 0}
\left(-\epsilon_1\epsilon_2 \log \langle a |I^{(N/2)}\rangle\right)$ is
the prepotential of the $(A_1,D_N)$ theory.
The parameter, $\zeta$, is a free parameter that can
be absorbed or emerged by rescaling the dynamical scale $\Lambda$.

Given the expression \eqref{eq:U2A1DN}, the factors \eqref{eq:AD1} and \eqref{eq:AD2} are interpreted as the contribution of
the $(A_1,D_{N})$ theories at the fixed point corresponding to
$(Y_1,Y_2)$ on the $U(2)$ instanton moduli space. Note that the gauge group is
now $U(2)$ instead of $SU(2)$, and the difference between \eqref{eq:AD1}
and \eqref{eq:AD2} is how the $U(1)\subset U(2)$ is coupled to the
$(A_1,D_N)$ theory. 

An advantage of the expression \eqref{eq:U2A1DN} is that one can easily
introduce an extra fundamental hypermultiplet by multiplying
$\mathcal{Z}^\text{fund}_{Y_1,Y_2}(a,M)$ to the summand, where $M$
is the mass of the hypermultiplet. In particular, setting $N=4$ in
\eqref{eq:U2A1DN} and introducing an extra fundamental hypermultiplet,
the partition function is now
\begin{align}
 \mathcal{Z}_{U(2)} = \mathcal{Z}_\text{pert}\sum_{Y_1,Y_2}q^{|Y_1|+|Y_2|}\mathcal{Z}_{Y_1,Y_2}^\text{vec}(a)\mathcal{Z}_{Y_1,Y_2}^{(A_1,D_4)}(a,b,u)\tilde{\mathcal{Z}}^{(A_1,D_4)}_{Y_1,Y_2}(a,\tilde{b},\tilde{u})\mathcal{Z}_{Y_1,Y_2}^\text{fund}(a,M)~,
\end{align}
where $\Lambda^{b_0}$ is now replaced by $q$ since the $SU(2)$ gauge
coupling is exactly marginal.
This is almost equivalent to the instanton partition function of the
$(A_3,A_3)$ theory. The only difference from the $(A_3,A_3)$ is that the
$SU(2)$ gauge group in Fig.~\ref{fig:quiver1} is replaced by
$U(2)$, which gives rise to an extra prefactor, $\mathcal{Z}_{U(1)}$, of
the partition function. Therefore, the partition function of the
$(A_3,A_3)$ theory is evaluated as
\begin{align}
 \mathcal{Z}_{(A_3,A_3)} = \frac{\mathcal{Z}_{U(2)}}{\mathcal{Z}_{U(1)}}~.
\end{align}

\section{$U(2)$-version of generalized AGT for $(A_1,D_\text{odd})$}

\label{sec:U2}

In this section, we will extend the $U(2)$-version of the generalized
AGT correspondence reviewed in Sec.~\ref{sec:review} to the case of
$(A_1,D_N)$ theories for odd $N$. Specifically, we will generalize
\eqref{eq:AD1} to the case of odd $N$.\footnote{The same generalization
is possible for \eqref{eq:AD2}, but we will focus on generalizing
\eqref{eq:AD1} here to make our argument concise.}

Even when $N$ is odd, the $(A_1,D_{N})$ theory is still realized by
compactifying 6d (2,0) $A_1$ theory on sphere with an irregular and a
regular puncture. Therefore, exactly the same discussion as in
Sec.~\ref{subsec:GAGT} leads us to identifying
\begin{align}
 \mathcal{Z}_{(A_1,D_N)} = \langle a | I^{(N/2)}\rangle~,
\end{align}
as the partition function of the $(A_1,D_N)$ theory. 
From the equivalence between
\eqref{eq:curve0} and \eqref{eq:curve1}, we see that the non-vanishing
eigenvalues, $\lambda_N,\cdots,\lambda_{\frac{N+1}{2}}$, in
\eqref{eq:eigen1} appear as the coefficients of the first
$\frac{N-1}{2}$ non-trivial terms in the SW curve:\footnote{Here we
absorbed $\lambda_N$ in front of $z^{N-2}$ by rescaling $z$ and $x$ so that
$xdz$ is kept fixed. The fact that we can absorb $\lambda_N$ this way
reflects the conformal invariance of $(A_1,D_{N})$.}
\begin{align}
 x^2 &= \frac{1}{z^{N+2}} -
 \frac{\lambda_{N-1}}{(-\lambda_N)^{\frac{N-1}{N}}}\frac{1}{z^{N+1}} -
 \frac{\lambda_{N-2}}{(-\lambda_N)^{\frac{N-2}{N}}}\frac{1}{z^N} - \cdots -
 \frac{\lambda_{\frac{N+1}{2}}}{(-\lambda_N)^{\frac{N+1}{2N}}}\frac{1}{z^{\frac{N+5}{2}}}
 + \cdots~,
\label{eq:A1DN}
\end{align}
which are identified as the relevant couplings
of $(A_1,D_{N})$ for odd $N$ \cite{Cecotti:2010fi, Xie:2012hs}. Therefore the relevant couplings of $(A_1,D_N)$ theories
are all encoded in the eigenvalues of $L_{\frac{N+1}{2}},\cdots,L_{N-2}$
and $L_{N-1}$ (normalized by that of $L_N$). This is a straightforward generalization of what we reviewed
in Sec.~\ref{subsec:GAGT} to odd $N$.

One difficulty for odd $N$ is, however, the
irregular state $|I^{(N/2)}\rangle$ cannot be obtained in a colliding limit of regular
primary operators. As such, any result derived via the colliding limit for
even $N$ is not available for odd $N$. For instance, while $\lambda_k$
are translated into $c_k$ through \eqref{eq:lambda} for even $N$,
a similar translation is not available for odd $N$. As a result, an explicit expression for the action of
$L_1\cdots,L_{\frac{N-1}{2}}$ on $|I^{(N/2)}\rangle$ has not been identified for
odd $N$.

The lack of a colliding-limit construction gives rise to another difficulty when
considering the $U(2)$-version of the generalized AGT
correspondence. Generalizing the argument in Sec.~\ref{subsec:U2}, it is
natural to expect that there exists an irregular state $|\hat{I}^{(N/2)}\rangle$ of $Vir\oplus H$ such
that 
\begin{align}
 \mathcal{Z}_{Y_1,Y_2}^{(A_1,D_N)} \sim \frac{\langle
 a;Y_1,Y_2|\hat{I}^{(N/2)}\rangle}{\langle a|\hat{I}^{(N/2)}\rangle}~,
\label{eq:conj}
\end{align}
is identified, even for odd $N$, as the contribution from an $(A_1,D_N)$ sector at each
fixed point on the $U(2)$ instanton moduli space for the gauge theory
described by the quiver in Fig.~\ref{fig:quiver4}. Here, the irregular state
$|\hat{I}^{(N/2)}\rangle$ is decomposed as $|\hat{I}^{(N/2)}\rangle =
|I^{(N/2)}\rangle \otimes |I_H^{(N/2)}\rangle$, where
$|I^{(N/2)}\rangle$ is the irregular state of Virasoro algebra discussed
in the previous two paragraphs, and $|I^{(N/2)}_H\rangle$ is a rank-$\frac{N}{2}$
irregular state of Heisenberg algebra. For even $N$,
$|I^{(N/2)}_H\rangle$ is completely characterized by \eqref{eq:irreg-H},
which was derived via the colliding-limit construction of
$|I^{(N/2)}_H\rangle$. However, for odd $N$, the lack of a
colliding-limit construction makes it difficult to find a similar
characterization of $|I^{(N/2)}_H\rangle$. 

The above discussions imply that, due to the lack of colliding-limit, we do not know how $L_1,\cdots,
L_{\frac{N-1}{2}}$ and $a_{k>0}$ act on the irregular state
$|\hat{I}^{(N/2)}\rangle = |I^{(N/2)}\rangle \otimes
|I^{(N/2)}_H\rangle$ when $N$ is odd.
Without knowing these actions, one cannot compute  
\begin{align}
\frac{\langle a |a_{m_\ell}^{q_{\ell}}
\cdots a_{1}^{q_1} L_{n_k}^{p_k}\cdots L_{n_1}^{p_1}
 |\hat{I}^{(N/2)}\rangle}{\langle a|\hat{I}^{(N/2)}\rangle}~,
\label{eq:matrix-element}
\end{align}
for $n_i>0$ and $m_i>0$. This generically makes it hard
to compute \eqref{eq:conj} since $\langle a;Y_1,Y_2|$ is a linear
combination of vectors of the form $\langle a |a_{m_\ell}^{q_{\ell}}
\cdots a_{1}^{q_1} L_{n_k}^{p_k}\cdots L_{n_1}^{p_1}$. 
In the next four sub-sections, however, we will argue that this difficulty can be overcome
when we focus on the classical limit $\epsilon_1,\epsilon_2\to 0$ and
turn off relevant couplings and the VEVs of Coulomb branch operators in the $(A_1,D_{N})$-sector.

\subsection{Classical limit as the commutative limit}

While the irregular state $|I^{(N/2)}\rangle$ is an eigen state of
$L_{\frac{N+1}{2}},\cdots,L_N$ with non-vanishing eigenvalues, it is not an eigen state of
$L_1,\cdots,L_{\frac{N-1}{2}}$. Indeed, the Virasoro algebra 
\begin{align}
 [L_n,L_m] = (n-m)L_{n+m} + \frac{n(n^2-1)}{12}\delta_{n+m,0}
\label{eq:Vir}
\end{align}
forbids $L_{1},\cdots,L_N$ to have non-vanishing eigenvalues when
$N>2$. This is the
main reason that \eqref{eq:eigen2} (which is only for even $N$) involves differential operators on
the RHS for $0\leq k<\frac{N}{2}$.

However, when computing the matrix element \eqref{eq:matrix-element} in the classical limit $\epsilon_1,\epsilon_2\to
0$, the sub-algebra formed by $\{L_{n> 0}\}$ reduces to a commutative
algebra. The reason for this is the following. First, in the context of
the generalized AGT correspondence, the SW curve \eqref{eq:curve1} of a 4d theory is
identified as \eqref{eq:curve0} on the 2d side. This and the fact that the SW 1-form,
$xdz$, has scaling dimension $1$ imply that $z$ and $T(z)$ in \eqref{eq:curve0} has
four-dimensional scaling dimensions $\Delta_\text{4d}(z)=-2/N$ and
$\Delta_\text{4d}(T(z)) = \Delta_\text{4d}(x^2) = 2(1+2/N)$, respectively. Since the stress tensor is
expanded as
\begin{align}
 T(z) = \sum_{n\in\mathbb{Z}}\frac{L_n}{z^{n+2}}~,
\end{align}
this implies that, when acting on $|I^{(N/2)}\rangle$, $L_n$ is
associated with four-dimensional scaling dimension
\begin{align}
 \Delta_\text{4d}(L_n)= 2\left(1-\frac{n}{N}\right)~.
\label{eq:dimLn}
\end{align}
Recall here that, in the AGT correspondence, the 4d scaling dimensions
are invisible since we set $\epsilon_1\epsilon_2=1$, as explained around Eq.~(3.2) of \cite{Alday:2009aq}. To recover the
correct scaling dimensions, we need to multiply every quantity of
dimension $\Delta_\text{4d}$ by
$(\epsilon_1\epsilon_2)^{\Delta_\text{4d}/2}$. This particularly means
the replacement $L_n \to (\epsilon_1\epsilon_2)^{1-\frac{n}{N}}L_n$, and
therefore 
\begin{align}
[L_n,L_m] = (n-m)(\epsilon_1\epsilon_2)L_{n+m}~,
\end{align}
for $m,n>0$.
This implies that, when focusing on the leading term in
the limit $\epsilon_1,\epsilon_2\to0$, the sub-algebra formed
by $\{L_{n>0}\}$ reduces to a commutative algebra.
Therefore, in the computation of \eqref{eq:matrix-element} in the
classical limit, one can regard all $L_n$ and $a_m$ as
commutative and simultaneously diagonalizable.

This suggests the following conjecture: {\it in the classical limit $\epsilon_1,\epsilon_2\to 0$,
the irregular states $|I^{(N/2)}\rangle$ approaches to a simultaneous
eigen state of $\{L_{n>0}\}$ and $\{a_{m>0}\}$.} As seen in
\eqref{eq:eigen2}, this is indeed
the case when $N$ is even; in the third line of the RHS of \eqref{eq:eigen2},
$\sum_{\ell=1}^{N/2-k}\ell \,c_{\ell+k}\frac{\partial}{\partial c_\ell}$
is sub-leading in the classical limit, and therefore $|I^{(N/2)}\rangle$
approaches to a simultaneous eigen state of $\{L_k\}$ and $\{a_k\}$ in
the classical limit. We here assume that the above conjecture is
also satisfied for odd $N$. Then the matrix
element \eqref{eq:matrix-element} can be evaluated in the classical
limit as
\begin{align}
\frac{\langle a |a_{m_\ell}^{q_{\ell}}
\cdots a_{1}^{q_1} L_{n_k}^{p_k}\cdots L_{n_1}^{p_1}
 |\hat{I}^{(N/2)}\rangle}{\langle a|\hat{I}^{(N/2)}\rangle} \;=\; \left(\prod_{i=1}^\ell(\mathfrak{a}_{m_i})^{q_i}\right)\left(\prod_{j=1}^{k}(\mathfrak{b}_{n_j})^{p_j}\right)~,
\label{eq:reduced-matrix-element}
\end{align}
where $\mathfrak{a}_i$ and $\mathfrak{b}_i$ are defined by
\begin{align}
 \mathfrak{a}_m \equiv \frac{\langle a |
 a_m|\hat{I}^{(N/2)}\rangle}{\langle a|\hat{I}^{(N/2)}\rangle}~,\qquad
 \mathfrak{b}_n \equiv \frac{\langle
 a|L_n|\hat{I}^{(N/2)}\rangle}{\langle a|\hat{I}^{(N/2)}\rangle}~,
\label{eq:eigens}
\end{align}
for $m,n>0$.\footnote{The reduction of
\eqref{eq:matrix-element} to \eqref{eq:reduced-matrix-element} was
explicitly observed in the case of $N=4$ in Sec.~5.1 of \cite{Kimura:2020krd}.}
Note here that, from \eqref{eq:eigen1}, we see that $\mathfrak{b}_{n} =
0$ for $n>N$. Therefore, 
\eqref{eq:reduced-matrix-element} is a function of $\mathfrak{b}_n$ for $n=1,\cdots,N$ and $\mathfrak{a}_{m}$
for $m>0$. Note also that, for $\frac{N+1}{2}\leq n \leq N$,
$\mathfrak{b}_n$ is identical to the eigenvalues $\lambda_n$ in \eqref{eq:eigen1}.

\subsection{4d scaling dimensions of 2d parameters}

Here we evaluate the 4d scaling dimensions of the parameters
$\{\mathfrak{a}_m\}$ and $\{\mathfrak{b}_n\}$ defined above. We will use
them in the next sub-section to argue that, when all the couplings and VEVs
of Coulomb branch operators of $(A_1,D_N)$ are turned off, one has $\mathfrak{b}_n =
\mathfrak{a}_m=0$ for all $n\neq N$ and $m>0$.

To that end, we first see from \eqref{eq:dimLn} that
\begin{align}
 \Delta_\text{4d}\left(\mathfrak{b}_n\right) =
 2\left(1-\frac{n}{N}\right)~,
\label{eq:dimb}
\end{align}
which implies that $\mathfrak{b}_n$ for $n>N$ are of negative dimensions and
therefore irrelevant in the infrared. Since the Nekrasov partition
function is the quantity defined in the infrared,
\eqref{eq:conj} must be independent of such parameters. This is
consistent with
the condition $\mathfrak{b}_n =0$ for $n>N$.

Let us now turn to the scaling dimensions of $\mathfrak{a}_m$. To evaluate them, one needs to
use explicit expressions for the basis $|a;Y_1,Y_2\rangle$ of the
highest weight module of $Vir\oplus H$. As shown in \cite{Alba:2010qc}, the state $|a;Y_1,Y_2\rangle$ is generally a linear combination of descendants of the
highest weight state $|a\rangle$ of degree $(|Y_1|+|Y_2|)$. Here, the
degree is defined by the sum of the degrees in the sense of Virasoro and
Heisenberg algebras; for instance, the degree of
$(L_{-1})^2a_{-5}|a\rangle$ is evaluated as seven. A few
examples of $|a;Y_1,Y_2\rangle$ are shown below:
\begin{align}
 |a;\emptyset,\emptyset\rangle &= |a\rangle~,
\label{eq:basis1}
\\
|a;{\tiny \yng(1)},\emptyset\rangle &=
 \Big(-i\left(\epsilon_1+\epsilon_2+2a\right)a_{-1}-L_{-1}\Big)|a\rangle~,
\\
|a;{\tiny \yng(2)},\emptyset\rangle
 &=\Big(-i\epsilon_1(\epsilon_1+\epsilon_2+2a)(2\epsilon_1+\epsilon_2)a_{-2}-(\epsilon_1+\epsilon_2+2a)(2\epsilon_1+\epsilon_2+2a)a_{-1}^2~,
\nonumber\\
&\qquad \qquad +2i(2\epsilon_1+\epsilon_2+2a) a_{-1}L_{-1} -
 \epsilon_1(\epsilon_1+\epsilon_2+2a)L_{-2} + L_{-1}^2
 \Big)|a\rangle~,
\\
|a;{\tiny \yng(1)},{\tiny \yng(1)}\rangle
 &=\Big(-i(\epsilon_1+\epsilon_2)a_{-2}-
 (\epsilon_1^2+\epsilon_2^2+\epsilon_1\epsilon_2-4a^2)a^2_{-1}+2i(\epsilon_1+\epsilon_2)a_{-1}L_{-1}
 -L_{-2}+L_{-1}^2\Big)|a\rangle~,
\label{eq:basis4}
\end{align}
where we recovered the complete $\epsilon_i$-dependence.
In the context of the AGT correspondence, the highest weight
$a$ of $|a\rangle$ is identified as the mass of the W-boson that arises on
the Coulomb branch of $SU(2)$ gauge theory, and therefore has scaling
dimension one. Similarly the
$\Omega$-deformation parameters
$\epsilon_i$ have scaling dimension one, i.e.,
\begin{align}
 \Delta_\text{4d}(a) = \Delta_\text{4d}(\epsilon_1) =
 \Delta_\text{4d}(\epsilon_2)=1~.
\label{eq:dima}
\end{align}
Combining this with the expressions for
$|a;Y_1,Y_2\rangle$ shown in \eqref{eq:basis1}--\eqref{eq:basis4},
one can read off the 4d scaling dimensions of $\mathfrak{a}_m =
\langle a|a_m|\hat{I}^{(N/2)}\rangle/\langle a
|\hat{I}^{(N/2)}\rangle$.

For instance, we see from \eqref{eq:basis1} and \eqref{eq:eigens} that $\mathcal{Z}_{{\tiny
\yng(1)},\emptyset}^{(A_1,D_N)} \sim \langle a;{\tiny
\yng(1)},\emptyset|\hat{I}^{(N/2)}\rangle/\langle
a|\hat{I}^{(N/2)}\rangle$ is evaluated as\footnote{Here, we recall that $\langle a;Y_1,Y_2|$ is obtained by expanding it as
a linear combinations of $L_{-n_1}^{p_1}\cdots
L_{-n_k}^{p_k}a_{-m_1}^{q_1}\cdots a_{-m_\ell}^{q_\ell}|a\rangle$ and replacing each of these vectors with $\langle a |a_{m_\ell}^{q_{\ell}}
\cdots a_{1}^{q_1} L_{n_k}^{p_k}\cdots L_{n_1}^{p_1}$ with the expansion
coefficients kept fixed.}
\begin{align}
 \mathcal{Z}_{{\tiny \yng(1)},\emptyset}^{(A_1,D_N)} \sim
 -i(\epsilon_1+\epsilon_2+2a)\mathfrak{a}_1 - \mathfrak{b}_1~,
\label{eq:ex1}
\end{align}
Since the two terms in \eqref{eq:ex1} must have the same scaling dimensions,
we see that
\begin{align}
 \Delta_\text{4d}(\mathfrak{a}_1) = \Delta_\text{4d}(\mathfrak{b}_1)-1 =
 1- \frac{2}{N}~.
\end{align}
The same analysis for $\mathcal{Z}_{{\tiny
\yng(2)},\emptyset}^{(A_1,D_N)}$ implies 
\begin{align}
 \Delta_\text{4d}(\mathfrak{a}_{2}) = 2\Delta_\text{4d}(\mathfrak{b}_1)
 -3 = 1- \frac{4}{N}~.
\end{align}

It is straightforward to do the same analysis for
all $\mathcal{Z}_{Y_1,\emptyset}^{(A_1,D_N)}$ with $Y_1= [1,\cdots,1]$. As
shown in \cite{Alba:2010qc}, the state $|a;Y_1,\emptyset\rangle$ is concisely
expressed as
\begin{align}
 |a;Y_1,\emptyset\rangle &= \Omega_{Y_1}(a)\text{\bf
 J}_{Y_1}^{(-\epsilon_2^2)}(x)|a\rangle~,
\label{eq:left}
\end{align}
where $\Omega_{Y_1}(a) \equiv (-\epsilon_1)^{|Y_1|}\prod_{(j,k)\in Y_1}(2a +j\epsilon_1+k\epsilon_2)$, and $\text{\bf J}_{Y_1}^{(1/g)}(x)$ is the
normalized Jack polynomial of variables $x\equiv
(x_1,x_2,\cdots)$.\footnote{Note that we have $g=-\epsilon_2^2$ here.} 
Here, the variables $(x_1,x_2,\cdots)$ are related to the $\{L_n\}$ and
$\{a_m\}$ as follows. First, write the Virasoro generators $L_{n\neq 0}$ as
\begin{align}
 L_n = \sum_{k\neq 0,n}c_kc_{k-n} + i(nQ -2a)c_n
\end{align}
in terms of $\{c_k\}$ such that $[c_k,c_\ell] =
\frac{k}{2}\delta_{k+\ell,0}$. Then $x=(x_1,x_2,\cdots)$ is related to $\{c_k\}$
and $\{a_m\}$ by the identifications
\begin{align}
 a_{-n} - c_{-n} = -i\epsilon_1 p_n(x)~,
\end{align}
where $p_n(x) \equiv \sum_{i=1}^{|Y_1|}x_i^n$. Therefore, to express
\eqref{eq:left} in terms of $\{a_m\}$ and $\{L_m\}$, one first needs to write
$\text{\bf J}_{Y_1}^{(-\epsilon_2^2)}(x)$ in terms of $\{p_n(x)\}$, and
then replace $p_n(x)$ with $i(a_{-n}-c_{-n})/\epsilon_1$.
When $Y_1=[1,\cdots,1]$, the Jack polynomial is simply ${\bf J}_{Y_1}^{(1/g)}(x)
= |Y_1|! \, \prod_{i=1}^{|Y_1|}x_i$. Rewriting this in terms of
$p_n(x) = i(a_{-n}-c_{-n})/\epsilon_1$ for $n\in\mathbb{N}$, one finds
that the expression \eqref{eq:left} for $Y_1= [1,\cdots,1]$ is of the form
\begin{align}
 |a;\,Y_1 = [1,\cdots,1],\,\emptyset\rangle &=
 \Bigg(
\mathcal{N}(Y_1)\,
 \epsilon_1^{|Y_1|-1}\left(\prod_{j=1}^{|Y_1|}(2a+j\epsilon_1+\epsilon_2)\right)a_{-|Y_1|} 
\nonumber\\
& \qquad +\; \left(-L_{-1}\right)^{|Y_1|} \;+\;
 \cdots \Bigg)|a\rangle~,
\label{eq:special}
\end{align}
where $\mathcal{N}(Y_1)$ is a numerical factor independent of $\epsilon_1$ and $\epsilon_2$.
Note here that the presence of $(-L_{-1})^{|Y_1|}$ on the right-hand
side of \eqref{eq:special} is already stressed in \cite{Alba:2010qc}.
The expression \eqref{eq:special} implies that, for $Y_1=[1,\cdots,1]$,
\begin{align}
 \mathcal{Z}_{Y_1=[1,\cdots,1], \emptyset}^{(A_1,D_N)} \sim
 \mathcal{N}(Y_1)\,\epsilon_1^{|Y_1|-1}\prod_{j=1}^{|Y_1|}(2a+j\epsilon_1+\epsilon_2)\mathfrak{a}_{|Y_1|}
 + (-\mathfrak{b}_{1})^{|Y_1|} + \cdots~.
\end{align}
For the first two terms on the right-hand side to be of the same scaling
dimension, we must have
\begin{align}
 \Delta_\text{4d}(\mathfrak{a}_m) = m\Delta_\text{4d}(\mathfrak{b}_1) -2m+1
 = 1-\frac{2m}{N}~.
\label{eq:dima}
\end{align}

\subsection{Computation of matrix elements for odd $N$}

In the rest of this paper, we focus on the classical limit
$\epsilon_1,\epsilon_2\to 0$ so that
\eqref{eq:reduced-matrix-element} is valid. In this case, it is sufficient to identify
the values of \eqref{eq:eigens} for the computation of \eqref{eq:conj}. 
Here we argue that, when the relevant couplings and VEVs of Coulomb
branch operators of $(A_1,D_N)$ are all turned off, the only non-vanishing parameter among
\eqref{eq:eigens} is $\mathfrak{b}_N$ and therefore
\eqref{eq:reduced-matrix-element} reduces to 
\begin{align}
\frac{\langle a |a_{m_\ell}^{q_{\ell}}
\cdots a_{1}^{q_1} L_{n_k}^{p_k}\cdots L_{n_1}^{p_1}
 |\hat{I}^{(N/2)}\rangle}{\langle a | \hat{I}^{(N/2)}\rangle} &=\left\{
\begin{array}{l} 
1 \qquad \text{for}\quad \ell=k=0
\\[2mm]
\delta_{n_1,N} (\mathfrak{b}_N)^{p_1} \qquad \text{for}\quad
\ell=0,\;\;k=1
\\[2mm]
0\qquad \text{for the others}\\
\end{array}
\right.~.
\label{eq:presc}
\end{align}

To derive \eqref{eq:presc}, we first note that all parameters of $(A_1,D_N)$ on the Coulomb branch are
encoded in the SW curve \eqref{eq:curve0}. Through the equivalence of
\eqref{eq:curve0} and \eqref{eq:curve1}, these are related to $a$ and the
non-vanishing components of $\mathfrak{b}_n$.
The interpretation of non-vanishing $\mathfrak{b}_n$ in four dimensions is as
follows. From \eqref{eq:dimb}, we see that
$\mathfrak{b}_1,\cdots,\mathfrak{b}_{\frac{N-1}{2}}$ are identified as
the VEVs of Coulomb branch operators since they have scaling dimensions
larger than one \cite{Argyres:1995jj, Argyres:1995xn, Eguchi:1996vu}. Similarly, $\mathfrak{b}_{\frac{N+1}{2}},\cdots,\mathfrak{b}_{N-1}$
are identified as relevant couplings since their dimensions are smaller
than one. Note that, the $(A_1,D_N)$ theory has no exactly
marginal coupling, and therefore the dimensionless parameter $\mathfrak{b}_N$
has no counterpart in four dimensions. This implies that the final result
must be independent of $\mathfrak{b}_N$, as discussed in the next sub-section.

Since the Coulomb branch of $(A_1,D_N)$ is completely
characterized by $\{\mathfrak{b}_n\}$ and $a$, any physical quantity of the $(A_1,D_N)$ theory (on the Coulomb
branch) should be determined by these parameters.
In particular, $\mathfrak{a}_m$ must be a function of $\{\mathfrak{b_n}\}$
and $a$. When $N$ is even, this function was
identified in \cite{Kimura:2020krd} via the colliding-limit construction of
$|\hat{I}^{(N/2)}\rangle$, where $\mathfrak{a}_m$ turned out to be
independent of $a$. Here we assume this independence to hold for odd
$N$ as well, and therefore $\mathfrak{a}_m$ is a function
only of
$\{\mathfrak{b}_n\}$.

While it is beyond the scope of this paper to compute
$\mathfrak{a}_m$ for generic values of $\{\mathfrak{b}_n\}$, one can easily compute it when all the relevant
couplings and VEVs of Coulomb branch operators are turned off in
the $(A_1,D_N)$ theory. Indeed, turning off these couplings and VEVs
implies that 
 \begin{align}
 \mathfrak{b}_n = 0~,\qquad \text{for}\qquad n\neq N~.
\end{align}
Note that this is equivalent to the condition that $\mathfrak{b}_n=0$
unless $\Delta_\text{4d}(\mathfrak{b}_n) = 0$. Since $\mathfrak{a}_m$ is
assumed to be
a function only of $\{\mathfrak{b}_n\}$, this implies that $\mathfrak{a}_m =
0$ unless $\Delta_\text{4d}(\mathfrak{a}_m) = 0$.\footnote{Note here that, since we are already taking the classical
limit $\epsilon_1,\epsilon_2\to 0$, the only non-vanishing dimensionful
parameter in the $(A_1,D_N)$ sector is now $a$. Since
$\mathfrak{a}_m$ is assumed to be independent of $a$, we see that $\mathfrak{a}_m =
0$ unless $\Delta_\text{4d}(\mathfrak{a}_m)=0$.}
From \eqref{eq:dima}, we see
that $\Delta_{\text{4d}}(\mathfrak{a}_m) = 0$ occurs if and only
if $m=N/2$, but this condition is never satisfied for odd $N$. Hence, we conclude that
\begin{align}
 \mathfrak{a}_m = 0~,
\end{align}
for all $m$, when the relevant couplings and the VEVs of Coulomb
branch operators of $(A_1,D_N)$ are turned off.
The above discussion implies that the matrix element \eqref{eq:reduced-matrix-element}
reduces to \eqref{eq:presc} when focusing on the classical limit
$\epsilon_1,\epsilon_2\to 0$ and turning off all the relevant couplings
and VEVs of Coulomb branch operators of the
$(A_1,D_N)$ theory. 

\subsection{Removing an unphysical degree of freedom}

Suppose that we turn off all the relevant couplings and VEVs of Coulomb
branch operators in $(A_1,D_N)$. Then one can compute the RHS of
\eqref{eq:conj} using \eqref{eq:conj}
and \eqref{eq:presc}. From \eqref{eq:presc}, we
 see that the result depends on $\mathfrak{b}_N$. 

Note that \eqref{eq:dimb} implies
$\Delta_\text{4d}\left(\mathfrak{b}_N\right)=0$, and therefore
$\mathfrak{b}_N$ must be an exactly marginal coupling if it is a
physical degree of freedom. However, the $(A_1,D_N)$ theory has no such coupling. This
means that $\mathfrak{b}_N$, that appears on the RHS of \eqref{eq:dimb},
is not a physical parameter in four dimensions. The fact that $\mathfrak{b}_N$ is unphysical can also be seen in the SW curve \eqref{eq:A1DN} of the $(A_1,D_N)$
theory; $\lambda_N = \mathfrak{b}_N$ can be absorbed by a change of
variables.
Hence, to make the relation \eqref{eq:dimb} more
precise, one has to introduce a prefactor on the RHS to remove this
unphysical degree of freedom.\footnote{This is exactly the same situation as in
\eqref{eq:AD1} for even $N$, where $(\zeta
c_{N/2})^{-\frac{2(|Y_1|+|Y_2|)}{N}}$ removes a degree of freedom
that has no physical meaning in the corresponding four-dimensional
theory.}

 As shown in \cite{Alba:2010qc}, the basis $|a;Y_1,Y_2\rangle$ is a
descendant at level $|Y_1|+|Y_2|$.\footnote{Here, the level of a
descendant means the sum of the level of the Virasoro descendant and that
of a Heisenberg descendant. For instance, $L_{-1}a_{-3}|a\rangle$ is a
descendant at level four.} Combining this fact with \eqref{eq:presc}, we
find that $\langle a;Y_1,Y_2|\hat{I}^{(N/2)}\rangle/\langle
a|\hat{I}^{(N/2)}\rangle$ is proportional to
$(\mathfrak{b}_N)^{\frac{|Y_1|+|Y_2|}{N}}$. This means that the following
expression is independent of $\mathfrak{b}_N$:
\begin{align}
 \mathcal{Z}_{Y_1,Y_2}^{(A_1,D_{N})}(a) = \left(\xi \mathfrak{b}_{N}\right)^{-\frac{|Y_1|+|Y_2|}{N}}\frac{\langle
 a;Y_1,Y_2|\hat{I}^{(N/2)}\rangle}{\langle a | \hat{I}^{(N/2)}\rangle}~,\label{eq:AD3}
\end{align}
where $\xi$ is a numerical free parameter that can be absorbed or
emerged by rescaling the dynamical scale. 
We therefore identify
\eqref{eq:AD3} as the precise expression for the contribution from
$(A_1,D_N)$ to the instanton partition function. Note that this is the
``odd-$N$ version'' of \eqref{eq:AD1}.
We will apply the above formula in the next
section to the computation of the instanton partition function of the $(A_2,A_5)$ theory.

\section{Application to the $(A_2,A_5)$ theory}
\label{sec:A2A5}

In this section, we compute the instanton partition function of  the
$(A_2,D_5)$ theory using our method described in the previous section.

\subsection{Partition function}

Recall that the $(A_2,A_5)$ theory is $SU(2)$ gauge theory 
described by the quiver diagram in Fig.~\ref{fig:quiver2}.
We first replace the gauge group with $U(2)$, and then the partition function of
the theory is evaluated as
\begin{align}
    \mathcal{Z}_{U(2)} = \mathcal{Z}^{U(2)}_\text{pert}\sum_{Y_1,Y_2}q^{|Y_1|+|Y_2|}
    \mathcal{Z}_{Y_1,Y_2}^\text{vec}(a)\mathcal{Z}_{Y_1,Y_2}^\text{fund}(a,M)\mathcal{Z}_{Y_1,Y_2}^{(A_1,D_3)}(a,d,u)
    \mathcal{Z}_{Y_1,Y_2}^{(A_1,D_6)}(a,m,\pmb{d},\pmb{u})~.
   \label{eq:Z-A2A5}
   \end{align}
Here $\mathcal{Z}_{Y_1,Y_2}^\text{vec}$ and
$\mathcal{Z}_{Y_1,Y_2}^\text{fund}$ are contributions respectively from
the vector multiplet and fundamental hypermultiplet \cite{Nekrasov:2002qd,Nekrasov:2003rj}, which have simple product expressions \cite{Flume:2002az,
Bruzzo:2002xf, Fucito:2004gi} as reviewed in (A.1) and (A.3) of \cite{Kimura:2020krd}.
On the other hand, $\mathcal{Z}_{Y_1,Y_2}^{(A_1,D_3)}$ and
$\mathcal{Z}_{Y_1,Y_2}^{(A_1,D_6)}$ are contributions respectively from the
$(A_1,D_3)$ and $(A_1,D_6)$ sectors in Fig.~\ref{fig:quiver2}. Here, $q$
is the exponential of the exactly marginal gauge coupling, $d$ and $u$ are respectively the relevant coupling and VEV of Coulomb branch operator 
in the $(A_1,D_3)$ theory, and $m$, $\pmb{d}=(d_1,d_2)$ and
$\pmb{u}=(u_1,u_2)$ are respectively the mass parameter,
relevant couplings and VEVs of Coulomb branch operators in the
$(A_1,D_6)$ theory.
The scaling dimensions of these parameters are as follows:
\begin{align}
    [q]=0,\qquad [d_1]=\frac{1}{3}~,\qquad
 [d]=[d_2]=\frac{2}{3}~,\qquad [u]=[u_1]=\frac{4}{3}~, \qquad [u_2]=\frac{5}{3}~.
\end{align}

In the rest of this section, we set $d=u=0$ so that our formula
derived in the previous section is available.
Using \eqref{eq:AD3}, we identify the contribution of the $(A_1,D_3)$ theory as
\begin{align}
      \mathcal{Z}_{Y_1,Y_2}^{(A_1,D_3)}(a)=\left(\xi\mathfrak{b}_3\right)^{-\frac{|Y_1|+|Y_2|}{3}}\frac{\langle
 a;Y_1,Y_2|\hat{I}^{\left(3/2\right)}\rangle}{\langle
 a|\hat{I}^{\left(3/2\right)}\rangle}~.
\label{eq:piece1}
\end{align}
Since we turn off the relevant coupling and the VEV of the Coulomb
branch operator, the RHS of \eqref{eq:piece1} can be computed via
\eqref{eq:presc}.

The contribution of the $(A_1,D_6)$
theory was already identified in \cite{Kimura:2020krd} and have reviewed
in \eqref{eq:AD1}; substituting $N=6$ we find 
\begin{align}
    \mathcal{Z}_{Y_1,Y_2}^{(A_1,D_6)}(a,m,\pmb{d},\pmb{u})=\left(\zeta c_3\right)^{-\frac{|Y_1|+|Y_2|}{3}}\frac{\langle
 a;Y_1,Y_2|\hat{I}^{(3)}\rangle}{\langle a|\hat{I}^{(3)}\rangle}~,
\label{eq:piece2}
\end{align}
where $\pmb{d}=(d_1,d_2)$ and $\pmb{u}=(u_1,u_2)$ are identified as in
\eqref{eq:dk} and \eqref{eq:uk}. 
We choose the free parameter $\zeta$ to be $\zeta = 2/\xi$ so that the
expressions in the next sub-section are simple. Changing the value of
$\zeta$ or $\xi$ just corresponds to rescaling $q$.
The RHS of \eqref{eq:piece2} can be evaluated by using $|\hat{I}^{(3)}\rangle = |I^{(3)}\rangle \otimes
|I^{(3)}_H\rangle$ and the following equations:
\begin{align}
L_k|I^{(3)}\rangle &= 0\quad \text{for}\quad k\geq 7~,
 \\
    L_6|I^{(3)}\rangle &= -c_3^2|I^{(3)}\rangle~,
    \label{eq:L6}
    \\
    L_5|I^{(3)}\rangle &= -2c_2c_3|I^{(3)}\rangle~,
    \\
    L_4|I^{(3)}\rangle &= -\left(c_2^2+2c_3c_1\right)|I^{(3)}\rangle~,
    \\
    L_3|I^{(3)}\rangle &= -2\left(c_1c_2+c_3(c_0-2Q)\right)|I^{(3)}\rangle~,
    \\
    L_2|I^{(3)}\rangle &=
    \left(c_3\frac{\partial}{\partial c_1}-c_2(2c_0-3Q)-c_1^2\right)|I^{(3)}\rangle~,
    \label{eq:L2}
    \\
    L_1|I^{(3)}\rangle &= \left(2c_3\frac{\partial}{\partial c_2}+c_2\frac{\partial}{\partial c_1}
    -2c_1(c_0-Q)\right)|I^{(3)}\rangle~,
\label{eq:L0}
\end{align}
and
\begin{align}
a_k|I_H^{\left(3\right)}\rangle =
 \left\{
\begin{array}{l}
-ic_k|I_H^{\left(3\right)}\rangle \quad \text{for} \quad k=1,2,3
\\[2mm]
0 \quad \text{for} \quad k>3
\\
\end{array}~.\right.
\label{eq:ak}
\end{align}
Using \eqref{eq:Z-A2A5}, \eqref{eq:piece1} and \eqref{eq:piece2},
one can evaluate $\mathcal{Z}_{U(2)}/\mathcal{Z}_\text{pert}^{U(2)}$ order by order in $q$.

Recall that we have replaced the $SU(2)$ gauge group in
Fig.~\ref{fig:quiver2} with $U(2)$. This induces an extra prefactor of
the partition function, $\mathcal{Z}_{U(1)}$, that is called the
``$U(1)$-factor.'' The partition function of the original $(A_2,A_5)$
theory is then recovered by removing $\mathcal{Z}_{U(1)}$ from
$\mathcal{Z}_{U(2)}$, i.e.,
\begin{align}
 \mathcal{Z}_{(A_2,A_5)} = \frac{\mathcal{Z}_{U(2)}}{\mathcal{Z}_{U(1)}}~.
\end{align}
 Since $a$ is the VEV of a scalar field
in the $SU(2)$ vector multiplet, $a$ is neutral under $U(1)$. Therefore
we expect that $\mathcal{Z}_{U(1)}$ is 
independent of $a$. This means that, up to an $a$-independent prefactor,
$\mathcal{Z}_{U(2)}$ and $\mathcal{Z}_{(A_2,A_5)}$ are identical.

\subsection{S-duality from the prepotential relation}

\label{subsec:prepotential}

We here focus on the prepotential of the $(A_2,A_5)$ theory:
\begin{align}
 \mathcal{F}^{(A_2,A_5)} &\equiv \lim_{\epsilon_i\to 0}
 \left(-\epsilon_1\epsilon_2\log \mathcal{Z}_{(A_2,A_5)}\right)~.
    \label{eq:F}
\end{align}
Up to the $a$-independent term $\lim_{\epsilon_i\to
0}(-\epsilon_1\epsilon_2 \log \mathcal{Z}_{U(1)})$, this is identical to 
\begin{align}
 \lim_{\epsilon_i\to
 0}\left(-\epsilon_1\epsilon_2\log\mathcal{Z}_{U(2)}\right).
\end{align} 
The prepotential \eqref{eq:F} is generally decomposed into the perturbative and instanton parts as 
\begin{align}
    \mathcal{F}^{(A_2,A_5)}=\mathcal{F}^{(A_2,A_5)}_{\text{pert}} +\mathcal{F}^{(A_2,A_5)}_{\text{inst}}~.
\end{align}
Again, up to $a$-independent terms affected by the $U(1)$-factor, the instanton part
$\mathcal{F}_{inst}^{(A_2,A_5)}$ is identical to
\begin{align}
 \lim_{\epsilon_i\to 0}\left(-\epsilon_1\epsilon_2\log
 \frac{\mathcal{Z}_{U(2)}}{\mathcal{Z}_\text{pert}^{U(2)}}\right)~,
\label{eq:F2}
\end{align}
which one can compute using the formula \eqref{eq:Z-A2A5}.\footnote{Note here that we are setting $d=u=0$ in \eqref{eq:Z-A2A5}, and therefore
\eqref{eq:F2} can be unambiguously computed via
\eqref{eq:Z-A2A5} with \eqref{eq:piece1} and \eqref{eq:piece2}.}
Below, we will compute this instanton part, and read off from it how the 
S-duality group acts on the UV gauge coupling of the $(A_2,A_5)$ theory.

To study the S-duality of the theory, it is useful to turn off the
couplings and VEVs in the
$(A_1,D_6)$ sector as well, i.e., $\pmb{d}=(0,0)$ and $\pmb{u}=(0,0)$ in
\eqref{eq:Z-A2A5}. 
In this case, $\mathcal{F}^{(A_2,A_5)}_\text{inst}$ is a function
of $q, a$ and two mass parameters $M$ and $m$. Using \eqref{eq:piece1}
and \eqref{eq:piece2}, one obtains
\begin{align}
    \mathcal{F}_\text{inst}^{(A_2,A_5)}(q;a,m,M) &\sim
    \frac{1}{6}\left(a^2 + \frac{mM^3}{2}a^{-2}\right)q^3 \notag
    \\
    &\qquad + \frac{1}{192} \Bigg[13a^2 + \left(\frac{3}{4}m^2 M^2 + 8 mM^3 + 3 M^4\right)a^{-2} \notag
    \\
    &\qquad  - \left(\frac{9}{4}m^2 M^4 + 3 M^6\right)a^{-4} + \frac{5}{4}m^2M^6 a^{-6}\Bigg]q^6 + \mathcal{O}(q^9)~,
    \label{eq:F-A2A5}
\end{align}
where ``$\sim$'' means that the LHS and RHS are identical up to $a$-independent terms affected by the the $U(1)$-factor.
Remarkably, the above expression is in a striking resemblance to the
instanton part $\mathcal{F}_\text{inst}^{N_f=4}$ of the prepotential of $SU(2)$ gauge theory with four fundamental
flavors. Indeed, comparing \eqref{eq:F-A2A5} with \eqref{eq:F-Nf=4} in
appendix \ref{app:Nf=4}, we see that the relation
\begin{align}
    3\mathcal{F}_\text{inst}^{(A_2,A_5)}\left(q;a,m,M\right) =
    \mathcal{F}_\text{inst}^{N_f=4}\left(q^3;a,\frac{m}{2},M,M,M\right)
    \label{eq:relation-M}
\end{align}
holds, at least up to $\mathcal{O}(q^9)$!\footnote{To be precise, we
have only checked this relation up to $\mathcal{O}(q^9)$, and also up to terms affected by the $U(1)$-factor. }
Note that one of the four mass parameters on the RHS is
related to the mass parameter $m$ in the $(A_1,D_6)$ sector on the
LHS, while the remaining
three masses on the RHS are identified with the mass $M$ of the single
fundamental hypermultiplet on the LHS.
In the next sub-section, we will show that these mass relations
are consistent with the SW curves of $(A_2,A_5)$ and $SU(2)$
gauge theory with four flavors.

In the same spirit as \cite{Kimura:2020krd},
we 
conjecture that the relation \eqref{eq:relation-M} extends
to the full prepotential. This
particularly implies
that, when the mass parameters are also turned off, one finds
\begin{align}
    3\mathcal{F}^{(A_2,A_5)}(q;a) =
    \mathcal{F}^{N_f=4}(q^3,a)~.
\label{eq:relation}
\end{align}
This prepotential relation is extremely powerful since one
can study the S-duality of the $(A_2,A_5)$ theory via that of $SU(2)$
gauge theory with four flavors. 
To see this, first note that the prepotentials of the two theories must
be written as
\begin{align}
 \mathcal{F}^{(A_2,A_5)}(q;a) = \left(\log q_\text{IR}\right)a^2~,\qquad
 \mathcal{F}^{N_f=4}(q;a) = \left(\log \tilde{q}_\text{IR}\right)a^2~,
\label{eq:FqIR}
\end{align}
for dimensional reasons, where $q_\text{IR}$ and $\tilde{q}_\text{IR}$
are functions of the UV gauge coupling $q$. One can regard $q_\text{IR}$
and $\tilde{q}_\text{IR}$ as IR gauge couplings of these theories on the
Coulomb branch. Indeed, in the weak coupling limit, both $q_\text{IR}$ and
$\tilde{q}_\text{IR}$ coincide with the UV gauge coupling $q$.

For $SU(2)$ gauge theory
with four flavors, the IR and UV gauge couplings are known to be related
by \eqref{eq:UV-IR-Nf=4} in appendix \ref{app:Nf=4} \cite{Grimm:2007tm}. This
theory is known to be invariant under an action of
$PSL(2,\mathbb{Z})$. Its action on the IR gauge coupling is written as
\begin{align}
T: \tilde{\tau}_\text{IR} \to \tilde{\tau}_\text{IR} + 1~,\qquad
 S:\tilde{\tau}_\text{IR}  \to -\frac{1}{\tilde{\tau}_\text{IR}}~,
\label{eq:TS1}
\end{align}
where $\tilde{\tau}_\text{IR} \equiv \frac{1}{\pi i}\log \tilde{q}_\text{IR}$. Through
\eqref{eq:UV-IR-Nf=4}, one can translate the above as
\begin{align}
 T: q\to \frac{q}{q-1}~,\qquad S: q\to 1-q~.
\label{eq:TS2}
\end{align}

Similarly, the $(A_2,A_5)$ theory is known to be invariant under
$PSL(2,\mathbb{Z})$ \cite{DelZotto:2015rca, Cecotti:2015hca, Buican:2015tda}. Indeed, the SW curve of the $(A_2,A_5)$ theory reduces to a
genus-one curve when dimensionful parameters except for $a$ are all
turned off. One difference from the previous paragraph is that 
the action of
$PSL(2,\mathbb{Z})$ on $q$ has not been identified, since the relation between
$q$ and $q_\text{IR}$ has been unclear for $(A_2,A_5)$. However, from the prepotential relation
we found above, one can now identify the
explicit relation between $q$ and $q_\text{IR}$ for the $(A_2,A_5)$
theory. Specifically, we see from \eqref{eq:relation} that
$\mathcal{F}^{(A_2,A_5)}(q;a)$ is obtained from
$\mathcal{F}^{N_f=4}(q;a)$ by the replacement
\begin{align}
q \longrightarrow q^3~,\qquad  \tilde{q}_\text{IR}  \longrightarrow
 q_\text{IR}^3~.
\label{eq:replacement}
\end{align}
Applying this replacement to \eqref{eq:UV-IR-Nf=4}, we find the
following relation between the UV and IR gauge couplings of the
$(A_2,A_5)$ theory:
\begin{align}
    q^3=\frac{\theta_2(q_\text{IR}^3)^4}{\theta_3(q_\text{IR}^3)^4}~.
\label{eq:UV-IR-A2A5}
\end{align}
This suggests that the $PSL(2,\mathbb{Z})$ acts on the IR gauge coupling
$\tau_\text{IR}\equiv \frac{3}{\pi}\log q_\text{IR}$
as
\begin{align}
    T:\tau_\text{IR} \to \tau_\text{IR} + 1~,\qquad S:\tau_\text{IR} \to
    -\frac{1}{\tau_\text{IR}}~,
    \label{eq:TS3/2}
\end{align}
 and on the UV gauge coupling as
\begin{align}
    T: q^3\to\frac{q^3}{q^3-1}~,\quad S: q^3\to 1-q^3~. 
\label{eq:TS3/2-2}
\end{align}
Indeed, applying \eqref{eq:replacement} to \eqref{eq:TS1} and
\eqref{eq:TS2}, one obtains \eqref{eq:TS3/2} and \eqref{eq:TS3/2-2}.

Remarkably, the above $PSL(2,\mathbb{Z})$-action on the $(A_2,A_5)$ theory
can be extended to a more non-trivial situation. Let us now turn on
$u_1$ and $u_2$ in \eqref{eq:Z-A2A5} while keeping $d,d_1,d_2$ and $u$ vanishing. Then
the resulting $\mathcal{F}_\text{inst}^{(A_2,A_5)}$ is a function of
$a,m,M,u_1,u_2$ and $q$. We find that this
$\mathcal{F}_\text{inst}^{(A_2,A_5)}$ is invariant under the following
change of variables:
 \begin{align}
    q&\to \frac{e^{\frac{\pi i}{3}}q}{(1-q^3)^{\frac{1}{3}}}~,\qquad 
    m  \to -m~,\qquad u_1 \to e^{\frac{2\pi i}{3}}u_1~,\qquad u_2 \to e^{\frac{\pi i}{3}}u_2~,
    \label{eq:T-trans}
\end{align}
where $M$ and $a$ are kept fixed. 
We checked this invariance up to $\mathcal{O}(q^{6
})$.
Note that the transformation \eqref{eq:T-trans} is a natural
extension of the $T$-transformation in \eqref{eq:TS3/2-2}. We believe
this can be further extended to the case of non-vanishing $d,d_1,d_2$ and
$u$. In particular, we believe the $T$-transformations for non-vanishing
$d,d_1$ and $d_2$ involve a non-trivial $q$-dependence as in the case of
$(A_3,A_3)$ theory studied in \cite{Kimura:2020krd}.
 We leave a careful study of it for future work.\footnote{As
discussed in Sec.~\ref{sec:U2},
 our formula for $\mathcal{Z}_{Y_1,Y_2}^{(A_1,D_3)}$ is
only for vanishing $d$ and $u$. Therefore, our
discussion on the S-duality here is limited to the case of $d=u=0$.
 Since $d$ and $d_2$
are of the same dimension, the $T$-transformation is expected to mix
them, which is why we turn off $d_2$ as well in the main text.
}

\section{Consistency with the Seiberg-Witten curve}
\label{sec:SW}

In this section, we show that the surprising relation
\eqref{eq:relation} is consistent with the SW curve of
the $(A_2,A_5)$ theory. In particular, we will show that the relation
between the two sets of mass parameters can also be seen in the SW curve.
We also show that 
the $T$-transformation \eqref{eq:T-trans} corresponds to a symmetry of
the curve.

The SW curve of the $(A_2,A_5)$ theory can be written as \cite{Cecotti:2010fi, Xie:2012hs}
\begin{align}
    0 = x^{3}&+z^{6}- \frac{1}{\mathsf{q}}x^{2}z^{2}-\mathsf{q}xz^{4}+c_{20}x^{2}+x(c_{11}z+c_{10}) \notag \\
    & +c_{05}z^{5}+c_{04}z^{4}+c_{03}z^{3}+c_{02}z^{2}+c_{01}z-c_{00}~,
    \label{eq:curve}
\end{align}
where $\mathsf{q}$ corresponds to the exactly marginal gauge coupling,
and is a non-trivial function of $q_{\text{IR}}$. 
The SW 1-form is given by $\lambda=xdz$.
Since the mass of a BPS state is given by $\oint \lambda$, the 1-form
$\lambda$ has scaling dimension one, which fixes the dimensions of the
parameters in \eqref{eq:curve} as
\begin{align}
    [x]=\frac{2}{3}~,\qquad [z]=\frac{1}{3}~,\qquad [c_{ij}]=2-\frac{2i+j}{3}~,\qquad [\mathsf{q}]=0~.
\end{align}
The coefficients $c_{ij}$ with $0<[c_{ij}]<1$ are regarded as relevant
couplings, while 
those with $[c_{ij}]>1$
are regarded as the VEV of Coulomb branch operators.
The remaining parameters, $c_{11}$ and $c_{03}$, are two mass parameters.

\subsection{Three sectors in the $(A_2,A_5)$ theory}

We first show that the curve \eqref{eq:curve} splits into three sectors in
the weak gauge coupling limit $\mathsf{q}\to 0$. 
To see this, let us study the behavior of the curve 
for $\mathsf{q}\sim 0$. As discussed in \cite{Buican:2014hfa},
the coefficients $c_{ij}$ of the curve must be renormalized so that as
many periods as possible are kept finite in the limit $\mathsf{q}\to
0$. We find that the correctly renormalized coefficients are as follows:
\begin{align}
    C_{ij}\equiv \mathsf{q}^{\frac{[c_{ij}]}{2}}c_{ij}\quad \text{for} \quad i\ne
 j~, 
\qquad    C_{11}\equiv \mathsf{q} c_{11}~,
\qquad C_{00}\equiv \mathsf{q} c_{00}~.
    \label{eq:normalize}
\end{align}
In terms of these renormalized parameters, the curve (\ref{eq:curve}) is
written as 
\begin{align}
    0 = x^{3}&+z^{6}- \frac{1}{\mathsf{q}}x^{2}z^{2}-\mathsf{q}xz^{4}+\mathsf{q}^{-\frac{1}{3}}C_{20}
    x^{2}+x(\mathsf{q}^{-1}C_{11}z+\mathsf{q}^{-\frac{2}{3}}C_{10}) \notag \\
  & +\mathsf{q}^{-\frac{1}{6}}C_{05}z^{5}+\mathsf{q}^{-\frac{1}{3}}C_{04}z^{4}
  +\mathsf{q}^{-\frac{1}{2}}C_{03}z^{3}+\mathsf{q}^{-\frac{2}{3}}C_{02}z^{2}
  +\mathsf{q}^{-\frac{5}{6}}C_{01}z^{1}-\mathsf{q}^{-1}C_{00}~.
  \label{eq:curve3}
\end{align}
One can show that the curve \eqref{eq:curve3} splits into the following
three sectors when we take $\mathsf{q}\to 0$ with $C_{ij}$ kept finite.
\begin{itemize}
    \item 
    In the region $|z/x|\sim \mathsf{q}^{-1/3}$, one has the curve 
    \begin{align}
        0=-\tilde{x}^{2}\tilde{z}^2+\tilde{z}^{6}+C_{11}\tilde{x}\tilde{z} + C_{05}\tilde{z}^{5}
        +C_{04}\tilde{z}^{4}+C_{03}\tilde{z}^3+C_{02}\tilde{z}^2+C_{01}\tilde{z}-C_{00}~, 
    \end{align}
    where we defined $\tilde{x}=\mathsf{q}^{-\frac{1}{6}}x$ and $\tilde{z}=\mathsf{q}^{\frac{1}{6}}z$.
One can shift $\tilde{x}$ as $\tilde{x}\to
	  \tilde{x}+C_{11}/(2\tilde{z})$ so that the curve coincides
	  with a known expression for the $(A_1,D_6)$ theory:
    \begin{align}
        \tilde{x}^{2}=\tilde{z}^{4}+C_{05}\tilde{z}^{3}
        +C_{04}\tilde{z}^{2}+C_{03}\tilde{z}+C_{02}+\frac{C_{01}}{\tilde{z}}-\frac{C_{00}-\frac{C_{11}^2}{4}}{\tilde{z}^{2}}~.
        \label{eq:A1D6curve}
    \end{align}
Note that the above shift of $\tilde{x}$ preserves the SW 1-form up to
	  exact terms. Here, we see that $C_{05}$ and $C_{04}$ are relevant
	  couplings, $C_{02}$ and $C_{01}$ are the VEVs of Coulomb
	  branch operators, and $C_{03}$ and $\sqrt{C_{00} - C_{11}^2/4}$ are
	  mass parameters of the $(A_1,D_6)$ theory. In particular, $\sqrt{C_{00}-C_{11}^2/4}$ is associated with the $SU(2)$
	  flavor sub-group that is gauged by the $SU(2)$ vector
	  multiplet in Fig.~\ref{fig:quiver2}.
    \item 
In
the region $|z/x|\sim \mathsf{q}^{2/3}$, the curve reduces to 
    \begin{align}
        0= \tilde{x}^3-\tilde{x}^2\tilde{z}^2 + C_{20}\tilde{x}^2 + \tilde{x}(C_{11}\tilde{z}+C_{10}) - C_{00}~,
    \end{align}
where we defined 
$\tilde{x}=\mathsf{q}^{-\frac{1}{3}}x$ and
	  $\tilde{z}=\mathsf{q}^{\frac{1}{3}}z$.
    By shifting and rescaling
the coordinates, this curve 
is further rewritten as
    \begin{align}
        0=X^{2}+Z^{4}+2^{\frac{1}{3}}C_{20}Z^{2}+4\sqrt{C_{00}-\frac{C_{11}^{2}}{4}}Z-2^{\frac{2}{3}}\left(C_{10}-\frac{C_{20}^{2}}{4}\right)~,
        \label{eq:A1D3curve}
    \end{align}
where we defined $X\equiv 2^{\frac{1}{3}} i \big(\tilde{x}+\frac{1}{2}(\tilde{z}^2 -
	  C_{20})\big)$ and $Z\equiv -2^{-\frac{1}{3}} i \tilde{z}$.
    We note that 
this coincides with the curve of the $(A_1,D_3)$ theory. In particular,
	  $C_{20}$ is the relevant coupling, $(C_{10}-C_{20}^2/4)$ is
	  the VEV of the Coulomb branch operator, and
	  $\sqrt{C_{00}-C_{11}^2/4}$ is the mass parameter associated
	  with the $SU(2)$ flavor symmetry. 

    \item
    In the region $|z/x|\sim 1$, 
the curve reduces to 
    \begin{align}
        0 = -x^2 z^2 + C_{11}xz -C_{00}~,
    \end{align}
	 which describes a weak coupling limit of the $SU(2)$
	 superconformal QCD as discussed in \cite{Buican:2014hfa}. In
	 particular, $C_{11}$ is identified as the mass parameter of a
	 fundamental hypermultiplet.
\end{itemize}
As seen above, in the limit $\mathsf{q}\to 0$, the curve of the
$(A_2,A_5)$ theory splits into the curves of the three
sectors shown in Fig.~\ref{fig:quiver2}.
Moreover, we have seen physical meanings of $C_{ij}$ in these
three sectors, which leads to 
the following identification of parameters in
\eqref{eq:Z-A2A5} in terms of those in the SW curve
\eqref{eq:curve3}:\footnote{Here, numerical factors in front of $C_{03}$
and $C_{11}$ are not physical. They are introduced here just to avoid
unimportant numerical coefficients below.}
\begin{align}
    d_1=C_{05}~&,\qquad d_2=C_{04}~,\qquad m=-\frac{C_{03}}{6}~,\qquad u_1=C_{02}~,\qquad u_2=C_{01}~, \notag
    \\
    &d=C_{20}~,\qquad u=C_{10}-\frac{C_{20}^2}{4}~,\qquad M=-\frac{C_{11}}{12}~.
    \label{eq:identify}
\end{align}

\subsection{S-duality from the curve}

We now show that the $T$-transformation \eqref{eq:T-trans} that we
identified in Sec.~\ref{subsec:prepotential} corresponds to a symmetry
of the SW curve \eqref{eq:curve}. We first note that the curve \eqref{eq:curve} is invariant under the following transformation: 
\begin{align}
    &\mathsf{q}\to e^{\frac{2\pi i}{3}}\mathsf{q}~,\qquad c_{10} \to e^{-\frac{4\pi i}{9}}c_{10}~,\qquad c_{11} \to e^{-\frac{2\pi i}{3}}c_{11}~,\qquad c_{20} \to e^{-\frac{2\pi i}{9}}c_{20}~, \notag \\
    & c_{01} \to -e^{\frac{\pi i}{9}}c_{01}~,\qquad c_{02} \to -e^{-\frac{\pi i}{9}}c_{02}~,\qquad c_{03} \to e^{\frac{2\pi i}{3}}c_{03}~, \notag\\
    &c_{04} \to e^{\frac{4\pi i}{9}}c_{04}~,\qquad c_{05} \to e^{\frac{2\pi i}{9}}c_{05}~,\qquad c_{00} \to -e^{\frac{\pi i}{3}}c_{00}~.
    \label{eq:T-trans2}
\end{align}
\footnote{
    At the same time, we take the change of coordinates in the curve \eqref{eq:curve}
    \begin{align}
        (x,z) \to (e^{-\frac{2\pi i}{9}}x,e^{\frac{2\pi i}{9}}z)~.
    \end{align} 
}
In the weak coupling limit $\mathsf{q}\to 0$, one can translate the
above transformation into a transformation of parameters in the three sectors. Indeed, \eqref{eq:normalize}
and \eqref{eq:identify} imply that \eqref{eq:T-trans2} is equivalent to
\begin{align}
    \mathsf{q}\to e^{\frac{2\pi i}{3}}\mathsf{q}~,\qquad &d_{1}\to
 -e^{\frac{2\pi i}{3}}d_{1}~,\qquad d_{2}\to -e^{\frac{\pi i}{3}}d_{2}~,
 \qquad m\to -m~,\notag\\
    & u_{1}\to e^{\frac{2\pi i}{3}}u_{1}~,\qquad u_{2}\to e^{\frac{\pi i}{3}}u_{2}~,
\end{align}
in the weak coupling limit. Note that this is in perfect agreement with
our $T$-transformation \eqref{eq:T-trans}.\footnote{Recall
that we have set $d_2=0$ in Sec.~\ref{subsec:prepotential} and therefore
consistent with $d_2\to -e^{\frac{\pi i}{3}}d_2$.}
 This means that our $T$-transformation \eqref{eq:T-trans} corresponds
to a symmetry of the SW curve.

One can show that the above symmetry transformation \eqref{eq:T-trans2}
coincides with an S-duality transformation of the theory.
To see this, 
let us turn off $c_{ij}$ except for $c_{00}$.
In this case, the curve is written as
\begin{align}
    0=(x-\sqrt{\mathsf{q}}z^2)(x+\sqrt{\mathsf{q}}z^2)\left(x-\frac{z^2}{\mathsf{q}}\right)-c_{00}~.
    \label{eq:curve2}
\end{align}
By changing the coordinates,\footnote{
    In terms of $w=x/z^2$ and $v=z^3$, the curve (\ref{eq:curve2}) is written as 
    \begin{align}
        v^2=\frac{c_{00}}{(w^2-\mathsf{q})\left(w-\frac{1}{\mathsf{q}}\right)}
    \end{align}
    We consider the following 
change of variables:
    \begin{align}
        w\to \frac{w\mathsf{q}^{\frac{1}{2}}\sqrt{1+\sqrt{f}}+\mathsf{q}^{\frac{1}{2}}\sqrt{\frac{1-\sqrt{f}}{1+\sqrt{f}}}}{w\sqrt{1-\sqrt{f}}+1}~, \quad 
        v\to \frac{\sqrt{1+\sqrt{f}}}{2\mathsf{q}^{\frac{1}{2}}\sqrt{f}}v\left(w\sqrt{1-\sqrt{f}+1}\right)^2~,
    \end{align}
    where $f$ is defined by $f\equiv 1-\mathsf{q}^3$.
    The curve is now 
written as
    \begin{align}
        v^2=\frac{\tilde{u}}{(w^2+1)-fw^4}~, 
\label{eq:curve5}
    \end{align}
    where $\tilde{u}$ is defined by $\tilde{u}\equiv\frac{2(1-f)^\frac{1}{3}}{\sqrt{1+\sqrt{f}}}c_{00}$.
    The SW curve is now written as $\frac{1}{3}wdv$.
In terms of 
$\tilde{x}\equiv i\sqrt{\tilde{u}}w$ and
$y\equiv \tilde{u}^{\frac{3}{2}}/v$, the curve \eqref{eq:curve5} is expressed as \eqref{eq:4flavors}.
}
the curve is expressed as
\begin{align}
    y^2=(\tilde{x}^2-\tilde{u})^2-f\tilde{x}^4~,
    \label{eq:4flavors}
\end{align}
where $f$ is defined by $f\equiv 1-\mathsf{q}^3$ and 
the SW 1-form is now written as $\frac{i\tilde{u}}{3}\frac{d\tilde{x}}{y}$ up to exact terms.
This is a standard expression for the curve of SU(2) conformal QCD. As
discussed in \cite{Seiberg:1994aj, Argyres:1995wt}, there is an S-duality
transformation involving
\begin{align}
 \sqrt{1-f} \to -\sqrt{1-f}~, \qquad \tilde{u}\to \tilde{u}~,
\end{align}
which is equivalent in our case to 
\begin{align}
 \mathsf{q} \to e^{\frac{2\pi i}{3}}\mathsf{q}~,\qquad c_{00}\to
 -e^{\frac{\pi i}{3}}c_{00}~.
\end{align}
Since this is precisely the action of \eqref{eq:T-trans2} on
$\mathsf{q}$ and $c_{00}$, we conclude that our T-transformation
\eqref{eq:T-trans2} (or equivalently \eqref{eq:T-trans}) is an extension
of this S-duality transformation to the case of 
generic values of $c_{ij}$.

\subsection{Relation between mass parameters}

We have shown in \eqref{eq:relation-M} that the prepotential of
$(A_2,A_5)$ and that of $SU(2)$ gauge theory with four flavors are in a
surprising relation. In particular, one of the four mass parameters of
the latter is identified with the mass of the fundamental hypermultiplet
of the former, and the other three masses of the latter are identified
with the mass in the $(A_1,D_6)$ sector. In this sub-section, we
rederive this mass relation from the SW curve.

As seen above, the curve of the $(A_2,A_5)$ theory is identical to that
of $SU(2)$ conformal QCD when $c_{ij}=0$ except for
$c_{00}$. This can be generalized to the case of non-vanishing mass
parameters. When we turn on the two mass parameters $c_{03}$ and $c_{11}$, the curve
\eqref{eq:curve2} of the $(A_2,A_5)$ theory is slightly modified.
In terms of $w\equiv x/z^2$ and $v\equiv z^3$, the modified curve is
written as
\begin{align}
 0 &=
 v^2\left(w-\sqrt{\mathsf{q}}\right)(w+\sqrt{\mathsf{q}})\left(w-\frac{1}{\mathsf{q}}\right)
 + v\left(c_{03}+c_{11}w\right) - c_{00}~.
\end{align}
Defining $P_3(w)\equiv
(w-\sqrt{\mathsf{q}})(w+\sqrt{\mathsf{q}})(w-1/\mathsf{q})$ and shifting
$v$ as $v\to v-(c_{03}+c_{11}w)/(2P_3(w))$, we can rewrite the above as
\begin{align}
 v^2 &=
 \frac{c_{00}}{P_3(w)}
 + \frac{\left(c_{03}+c_{11}w\right)^2}{4P_3(w)^2}~,
\label{eq:curve6}
\end{align}
where the SW 1-form is now $\lambda = -\frac{1}{3}vdw$ up to exact
terms. 

We see that
\eqref{eq:curve6} is precisely of the same form as the mass-deformed
curve of $SU(2)$ conformal QCD with four flavors
\cite{Gaiotto:2009we}:
\begin{align}
 v^2 = \frac{U}{P_3(w)} + \frac{M_4(w)}{P_3(w)^2}~,
\label{eq:curve7}
\end{align}
where $U$ stands for a coordinate of the Coulomb branch, and $M_4(w)$
is a fourth-order polynomial of $w$ and related to the mass
parameters of the theory. Since there exists one constraint on the
coefficients of $M_4(w)$, there are four independent coefficients of
$M_4(w)$. These four independent degrees of freedom are encoded in the residues of the SW 1-form at
$w=\pm \sqrt{\mathsf{q}},\,1/\mathsf{q}$ and $\infty$. These residues
are known to be identified with the following linear combinations of the
mass parameters, $m_1,\cdots,m_4$, of fundamental hypermultiplets:
\begin{align}
 m_1\pm m_2~,\qquad m_3\pm m_4~.
\label{eq:residueNf=4}
\end{align}
Comparing \eqref{eq:curve6} and \eqref{eq:curve7}, we see that $(c_{03}
+ c_{11}w)^2$ in \eqref{eq:curve6} is identified with $M_4(w)$ in \eqref{eq:curve7}.
This implies that the four
mass parameters of the latter theory are related to
the two
mass parameters of the former.

To see more concretely the relation between the mass parameters, let us compute the residues
of the SW 1-form of the $(A_2,A_5)$ theory. From \eqref{eq:curve6}, we see that the residues of the 1-form
$\lambda = -\frac{1}{3}vdw$ at $w=\pm \sqrt{\mathsf{q}}, 1/\mathsf{q}$
and $\infty$
are respectively
\begin{align}
 -\frac{c_{03}\pm c_{11}\sqrt{\mathsf{q}}}{12 
 (\mathsf{q}-1/\sqrt{\mathsf{q}})}~,\qquad
-\frac{c_{03} + \frac{c_{11}}{\mathsf{q}}}{6\left(\frac{1}{\mathsf{q}} -
 \sqrt{\mathsf{q}}\right)\left(\frac{1}{\mathsf{q}}+\sqrt{\mathsf{q}}\right)}~,\qquad 0~,
\end{align}
which reduce in the weak-coupling limit $\mathsf{q}\to 0$ to
\begin{align}
\frac{m \pm 2M}{2}~,\qquad
 2M~,\qquad 0~.
\end{align}
We see that these residues coincide with \eqref{eq:residueNf=4} if we identify
\begin{align}
 m_1 = \frac{m}{2}~,\qquad m_2=m_3=m_4=M~.
\label{eq:mass-relation}
\end{align}
This implies that the mass-deformed SW curve of $(A_2,A_5)$ is
identical to that of the $SU(2)$ conformal QCD when the four
mass parameters of the latter are restricted as in
\eqref{eq:mass-relation}. Note here that the restriction
\eqref{eq:mass-relation} of mass parameters is 
precisely equivalent to the one observed in the relation \eqref{eq:relation-M} for
the prepotentials of these theories!\footnote{The coincidence of the
numerical factor $1/2$ in front of $m$ is a consequence of our
identification \eqref{eq:identify}, and therefore is not non-trivial. What
is non-trivial here is the coincidence that, both in
\eqref{eq:relation-M} and \eqref{eq:mass-relation}, three of the four mass parameters of
the $SU(2)$ conformal QCD are equal and proportional to $M$, and the
remaining one is proportional to $m$.} This is a very non-trivial
consistency check of \eqref{eq:F-A2A5} and our formula for
$\mathcal{Z}_{Y_1,Y_2}^{(A_1,D_3)}$ that we developed in Sec.~\ref{sec:U2}.

\section{Conclusion and Discussions}

In this paper, we have considered the $U(2)$-version of the generalized AGT
correspondence for $(A_1,D_N)$ theories for odd $N$, in terms of irregular states of the direct sum
of Virasoro and Heisenberg algebras $Vir\oplus H$. In contrast to the
$(A_1,D_\text{even})$ case, the action of $Vir\oplus H$ on the irregular
state cannot be obtained in a colliding limit of primary operators,
which makes it very difficult to compute the (normalized) inner product of the form
in \eqref{eq:conj}. However, we have shown that, when the relevant couplings and the VEVs of
Coulomb branch operators of the $(A_1,D_N)$ theory are turned off, one can
compute the inner product as in \eqref{eq:AD3}.

Using the formula \eqref{eq:AD3}, we have computed the instanton
partition function of the $(A_2,A_5)$ theory, i.e., the coupled system
of an $SU(2)$ vector multiplet, a fundamental hypermultiplet,
$(A_1,D_6)$ and $(A_1,D_3)$ as described in Fig.~\ref{fig:quiver2}. Our
result implies a surprising relation \eqref{eq:relation-M} between the
prepotential of the $(A_2,A_5)$ theory and that of the $SU(2)$
superconformal QCD. A similar relation was found in
\cite{Kimura:2020krd} for the $(A_3,A_3)$ theory. Using the relation
\eqref{eq:relation-M}, we have read off how the S-duality group acts on
parameters including the UV gauge coupling. We have also checked in
Sec.~\ref{sec:SW} that the relation \eqref{eq:relation-M} is consistent
with the Seiberg-Witten curves of the $(A_2,A_5)$ theory and the $SU(2)$
superconformal QCD.

One can also apply our formula for
$\mathcal{Z}_{Y_1,Y_2}^{(A_1,D_\text{odd})}$ to other gauged AD
theories. For instance, let us consider the $SU(2)$ gauge theory coupled to three copies of the $(A_1,D_3)$
theory. As in the case of $(A_2,A_5)$, the $SU(2)$ gauge coupling
of this theory is exactly marginal. Using our formula for
$\mathcal{Z}^{(A_1,D_3)}_{Y_1,Y_2}$, one can then compute the prepotential of
this theory, up to terms affected by the $U(1)$-factor, at least when the relevant
coupling and the VEV of Coulomb branch operator of the $(A_1,D_3)$
sectors are turned off. We have done this computation and checked up to
$\mathcal{O}(q^6)$ that the resulting prepotential has no instanton
correction at all.
Note that the same situation occurs for the prepotential of
$\mathcal{N}=4$ super Yang-Mills theories (SYMs). Indeed, a peculiar connection
between the $SU(2)$ gauge theory coupled to three $(A_1,D_3)$ theories and
$\mathcal{N}=4$\, $SU(2)$ SYM has already been pointed out in \cite{Buican:2020moo};
the Schur index of these two theories are related by a simple change of
variables. It would be very interesting to study this connection further.

There are clearly many future directions. One of the most important
directions is to understand the reason for the peculiar relation
\eqref{eq:relation-M} for the prepotentials.
 Another interesting
direction is to study the Nekrasov-Shatashvili limit of the instanton
partition function \cite{Nekrasov:2009rc}, which should be combined with
the recent results on the quantum periods of AD theories
\cite{Ito:2017iba, Ito:2018hwp, Ito:2019twh, Ito:2020lyu}. 
The uplift of
our formula \eqref{eq:AD3} to five dimensions would also be an
interesting direction. It would also be interesting to search for a
matrix model description of the instanton partition function of
$(A_2,A_5)$, generalizing the ones studied in
\cite{Nishinaka:2012kn, Grassi:2018spf, Itoyama:2018wbh, Itoyama:2018gnh, Itoyama:2021nmj, Oota:2021qky}.

\section*{Acknowledgements}

We are grateful to Matthew Buican, Kazunobu Maruyoshi, Jaewon Song, Yuji
Sugawara, Yuji Tachikawa and Takahiro Uetoko for discussions.
T.~N. is especially grateful to Matthew Buican for helpful discussions in
many collaborations on related topics. 
T.~N.'s research is partially supported by
JSPS KAKENHI Grant Numbers JP18K13547 and JP21H04993. This work was also
partly supported by Osaka Central Advanced Mathematical Institute: MEXT Joint Usage/Research Center on Mathematics and Theoretical Physics JPMXP0619217849.

\appendix

\section{S-duality of $SU(2)$ conformal QCD}
\label{app:Nf=4}

Here we give a brief review of an expression for the prepotential of
$SU(2)$ superconformal QCD, following \cite{Alday:2009aq}. 
When the mass parameters
are
 turned off, the prepotential must be written as
\begin{align}
    \mathcal{F}^{N_f=4}=(\log \tilde{q}_{\text{IR}}) a^2 ~,
\end{align}
for dimensional reasons, where $a$ is the VEV of the adjoint scalar in
the vector multiplet, and $\tilde{q}_\text{IR}$ is a function of the UV
gauge coupling $q$.
The 
above prepotential is written as the sum of the following perturbative 
and
instanton parts
\begin{align}
    \mathcal{F}^{N_f=4}_{\text{pert}}(a) =& (\log q - \log 16) a^2~,
    \\
    \mathcal{F}^{N_f=4}_{\text{inst}}(q;a) =& \left(\frac{1}{2}q + \frac{13}{64}q^2 + \frac{23}{192}q^3 + \cdots\right) ~,
\end{align}
from which the following relation between $q$ and $\tilde{q}_\text{IR}$ \cite{Grimm:2007tm}:
\begin{align}
    q =
 \frac{\theta_2\left(\tilde{q}_{\text{IR}}\right)^4}{\theta_3\left(\tilde{q}_{\text{IR}}\right)^4}~.
\label{eq:UV-IR-Nf=4}
\end{align}
This relation implies that there are S-dual transformations $S$ and $T$
such that
\begin{align}
    T: \tau_{\text{IR}}\to\ \tau_{\text{IR}}+1~, \qquad S:
 \tau_{\text{IR}} \to -\frac{1}{\tau_{\text{IR}}}~,
\label{eq:theta_Nf=4}
\end{align}
where $\tau_{IR}$ is defined by
\begin{align}
    \tau_{IR} \equiv \frac{1}{i\pi} \log \tilde{q}_{\text{IR}} = \frac{\theta_{\text{IR}}}{\pi} + \frac{8\pi i}{g_{\text{IR}}^2}~.
\end{align}
Note that 
the $T$-transformation corresponds to
 $\theta_{\text{IR}}\to \theta_{\text{IR}}+\pi$.
In terms of $q$, the above S-dual transformations are 
written as
\begin{align}
    T: q\to \frac{q}{q-1}~, \qquad S: q \to 1-q~.
    \label{eq:S-dual}
\end{align}

Let us now
turn on 
all the mass parameters. Then the instanton part of the prepotential is
modified as 
\begin{align}
    \mathcal{F}_\text{inst}^{N_f=4}&(q;a,m_i) \notag
    \\
    =&\frac{1}{2}(a^2+ m_1 m_2 m_3 m_4 a^{-2} )q \notag
    \\
    &+\frac{1}{64}\Big(13a^2 + (16m_1 m_2 m_3 m_4 + m_3^2 m_4^2 + m_2^2(m_3^2 + m_4^2)+ m_1^2(m_2^2 + m_3^2 + m_4^2))a^{-2} \notag
    \\
    &- 3(m_2^2 m_3^2 m_4^2 +m_1^2(m_3^2 m_4^2 + m_2^2 (m_3^2 + m_4^2)))a^{-4} + 5m_1^2 m_2^2 m_3^2 m_4^2 a^{-6} \Big)q^2+ \cdots ~.
   \label{eq:F-Nf=4}
\end{align}
The  S-dual transformations (\ref{eq:S-dual}) are 
now accompanied with the $SO(8)$ triality \cite{Seiberg:1994aj}.

\section{Decoupling a fundamental matter from $(A_2,A_5)$}

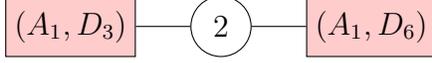
\begin{figure}
    \begin{center}
   \begin{tikzpicture}[gauge/.style={circle,draw=black,inner sep=0pt,minimum size=8mm},flavor/.style={rectangle,draw=black,inner sep=0pt,minimum size=8mm},AD/.style={rectangle,draw=black,fill=red!20,inner sep=0pt,minimum size=8mm},auto]

   \node[AD] (1) at (-2,0) {\;$(A_1,D_{3})$\;};
   \node[gauge] (2) at (0,0) [shape=circle] {\;$2$\;} edge (1);
   \node[AD] (3) at (2,0)  {\;$(A_1,D_{6})$\;} edge (2);

     \end{tikzpicture}
   \caption{The mass deformed theory of the $(A_2,A_5)$ theory}
   \label{fig:quiver3}
    \end{center}
   \end{figure}

Here, we consider the decoupling limit of the fundamental hypermultiplet
from the $(A_2,A_5)$ theory at the level of the SW curve. Recall that
the $(A_2,A_5)$ theory is described by the quiver in
Fig.~\ref{fig:quiver2}. When the fundamental hypermultiplet is decoupled,
the resulting theory is described by the quiver in
Fig.~\ref{fig:quiver3}. This theory is called the ``$\hat{A}_{3,6}$
theory'' in \cite{Bonelli:2011aa}.

To see this decoupling at the level of the SW curve, we take the mass of
the fundamental hypermultiplet to infinity, i.e., $C_{11} \to
\infty$ in \eqref{eq:curve3}.
For the periods of the curves to be finite, one needs to keep 
\begin{align}
    \Lambda\equiv -\frac{1}{2}\sqrt{\mathsf{q}}C_{11}
\label{eq:Lambda}
\end{align}
finite in this limit. The finite constant $\Lambda$ is then identified as a dynamical
scale of the resulting theory.
In terms of 
\begin{align}
    X\equiv -\frac{\Lambda}{\sqrt{\mathsf{q}}}\left(\frac{x}{z^2}\right)^{\frac{1}{3}}+z^3\left(\frac{x}{z^2}\right)^{\frac{2}{3}}~,\qquad
    Z\equiv \left(\frac{z^2}{x}\right)^{\frac{1}{3}}~,
\end{align}
the curve in the limit $C_{11}\to\infty$ 
is 
written as
\begin{align}
    X^2&=\Lambda^{5/3}C_{05}Z^3+\Lambda^{4/3}C_{04}Z^2+\Lambda C_{03}Z+\Lambda^{2/3}C_{02}+\frac{\Lambda^{1/3}C_{01}}{Z} \nonumber
    \\
    &\qquad -\frac{C_{00}+2\Lambda^2}{Z^2}+\frac{\Lambda^{2/3}C_{10}}{Z^3}+\frac{\Lambda^{4/3}C_{20}}{Z^4}+\frac{\Lambda^2}{Z^5}~,
\end{align}
and the SW 1-form is written as $\lambda=XdZ$ up to exact terms.
We see that 
the above curve is precisely identical to that of the $\hat{A}_{3,6}$
theory \cite{Bonelli:2011aa}.

Note that, by standard arguments, the gauge coupling of the
conformal theory on the Coulomb branch,
$\exp\left(i\theta_\text{IR}- \frac{8\pi^2}{g_\text{IR}^2}\right)$, is related to
the dynamical scale of the mass-deformed theory, $\Lambda$, by
\begin{align}
\frac{\Lambda}{C_{11}} = \exp\left(i\theta_\text{IR} - \frac{8\pi^2}{g_\text{IR}^2}\right)~.
\end{align}
Combining this with \eqref{eq:Lambda}, we see that
\begin{align} 
\mathsf{q} \propto \exp\left(2i\theta_\text{IR} - \frac{16\pi^2}{g_\text{IR}^2}\right)~.
\end{align}
Recall here that our $T$-transformation \eqref{eq:T-trans2} involves
$\mathsf{q}\to e^{\frac{2\pi i}{3}}\mathsf{q}$. Using the above
relation, one can translate this into
\begin{align}
 \theta_\text{IR} \to \theta_\text{IR} + \frac{\pi}{3}~,
\end{align}
which implies that the $T$-transformation exchanges the minimal magnetic
monopole with a dyonic particle whose electric charge is
$1/6$ of that of the W-boson, which is consistent with the fact that
$PSL(2,\mathbb{Z})$ naturally acts on a modified electro-magnetic charge
lattice \cite{Cecotti:2015hca}.

\bibliography{AGT}
\bibliographystyle{utphys}

\end{document}